\documentclass[12pt,a4paper,superscriptaddress,preprint,dvipdfmx]{revtex4}
\usepackage{graphicx}
\usepackage{amssymb,amsmath}
\usepackage{bm}
\usepackage{color}

\begin{document}

\title{Networks maximizing the consensus time of voter models}
\date{\today}
\author{Yuni Iwamasa}
\affiliation{Faculty of Engineering,
The University of Tokyo, 7-3-1, Hongo, Bunkyo-ku, Tokyo 113-8656, Japan}
\author{Naoki Masuda}
\email{naoki.masuda@brsitol.ac.uk}
\affiliation{Department of Mathematical Informatics,
The University of Tokyo, 7-3-1, Hongo, Bunkyo-ku, Tokyo 113-8656, Japan}
\affiliation{Department of Engineering Mathematics, University of Bristol,
Merchant Venturers Building,
Woodland Road, Clifton, Bristol BS8 1UB, United Kingdom}
\affiliation{CREST, JST, 4-1-8, Honcho, Kawaguchi, Saitama 332-0012, Japan}

\begin{abstract}
We explore the networks that yield the largest mean consensus time of voter models under different update rules. By analytical and numerical means, we show that the so-called lollipop graph, barbell graph, and double-star graph maximize the mean consensus time under the update rules called the link dynamics, voter model, and invasion process, respectively. For each update rule, the largest mean consensus time scales as $O(N^3)$, where $N$ is the number of nodes in the network.
\end{abstract}

\maketitle

\section{Introduction}

Collective opinion formation in society seems to be
an emerging phenomenon that occurs
in networks of agents that interact and exchange opinions.
The voter model is an individual-based stochastic model for collective
opinion formation
that has been studied in, above all, statistical physics and probability theory communities for decades \cite{Liggett1985book,Castellano2009RMP,Redner2001book,Krapivsky2010book}. 

In the voter model, a node assumes one of the different states (i.e.,
opinions) that can stochastically change over time. The configuration
in which all nodes take the same state, i.e., consensus, is the only
type of absorbing configuration of the voter model. The mean time to
consensus, which we denote by $\left<T\right>$, is a fundamental
property of the voter model. For a given size of the population,
$\left<T\right>$ depends on the network structure,
which suggests that opinions spread faster on some networks
than others. Not surprisingly, $\left<T\right>$ is small for
well-connected networks such as the complete graph for which $\left<T\right>=O(N)$, where $N$ is the
number of nodes in the network \cite{BenAvraham1990JPhysA}.
More complex networks with small mean path lengths
also yield linear or sublinear dependence of $\left<T\right>$
on $N$ \cite{Castellano2003EPL,Vilone2004PRE,Castellano2005AIPConf,Castellano2005PhysRevE,Sood2005PRL,Suchecki2005EPL,Suchecki2005PRE,Antal2006PRL,Sood2008PRE,Vazquez2008NewJPhys}.
In contrast, $\left<T\right>$ can be
large for networks in which communication between nodes is difficult
for topological reasons.  For example, the one-dimensional chain yields
$\left<T\right>=O(N^2)$
\cite{Cox1989AnnProb,Castellano2009RMP,Krapivsky2010book}.
Networks with community structure also yield $\left<T\right>=O(N^2)$ when
intercommunity links are rare \cite{Masuda2014arxiv_2cliques}
because coordination between the different
communities is a bottleneck of the entire process.

In this study, we explore network structure that maximizes the consensus time. In practice, answering this question may help us understand why consensus is difficult in real society \cite{Kuran1995book,Huckfeldt2004book}.
In theory, we expect that there exist networks for which $\left<T\right>$ is larger than $O(N^2)$ for the following reason. The mean consensus time of the voter model is often analyzed through the random walk. In fact, the so-called coalescence random walk is the dual process of the voter model, implying that the mean time before random walkers coalesce into one gives $\left<T\right>$
for the voter model \cite{Liggett1985book,Donnelly1983MPCPS,Durrett1988book,Cox1989AnnProb}. In addition, in the voter model on the chain,
positions of active links, i.e., boundaries between adjacent nodes possessing the opposite states, perform a random walk until different active links meet and annihilate. This relationship helps us to calculate
$\left<T\right>$ for the chain [i.e., $\left<T\right>=O(N^2)$]
through the hitting or cover time of the random walk.
Because the
hitting time
\cite{Mazo1982BellSysTechJ,Lawler1986DiscMath,Brightwell1990RandStructAlgor,Coppersmith1993SiamJDiscMath}
and cover time \cite{Chandra1989STOC}
of the random walk are known to scale as $O(N^3)$ for some networks, $\left<T\right>$ 
of the voter model may also scale as $O(N^3)$ for these and other networks.

In networks in which the degree (i.e., number of neighboring nodes for a node)
is heterogeneous,
the consensus time depends on the specific rule with which
we update the state of nodes \cite{Castellano2005AIPConf,Antal2006PRL,Sood2008PRE}.
Therefore, we explore the networks that maximize $\left<T\right>$ separately for different update rules. For each update rule, we
determine such networks by combining exact numerical calculations of $\left<T\right>$ for small networks (Sec.~\ref{sec:exact}), coarse analytical arguments
to evaluate lower and upper bounds of $\left<T\right>$
(Secs.~\ref{sub:order lollipop} and \ref{sub:order barbell}),
analysis of the coalescing random walk
(Sec.~\ref{sub:order double-star}), and
direct numerical simulations for large networks (Sec.~\ref{sec:direct numerical}). The results are summarized in Table~\ref{tab:summary}.

\section{Model}

We consider undirected networks possessing $N$ nodes
and $M$ links.
Each node possesses either of the two states $\mathbf{0}$ and $\mathbf{1}$ at any time $t$. In each update event,
the state of a node is updated, and time $1/N$ is consumed. We repeat this procedure until either the consensus of state $\mathbf{0}$ or that of state $\mathbf{1}$ is reached. It should be noted that
a node is updated once per unit time on average.

We examine three update rules according to Refs.~\cite{Antal2006PRL,Sood2008PRE}.
Under the link dynamics (LD) \cite{Castellano2005AIPConf,Suchecki2005EPL,Antal2006PRL,Sood2008PRE}, we first select a link with probability $1/M$. Denote by $i$ and $j$ the two endpoints of the selected link. With probability $1/2$, $i$ copies $j$'s state. With the remaining probability $1/2$, $j$ copies $i$'s state. In this manner, local consensus is obtained in a single update event. If $i$ and $j$ have the same state beforehand, the state of neither node changes. Under the voter model (VM) update rule \cite{Castellano2003EPL,Vilone2004PRE,Castellano2005PhysRevE,Sood2005PRL,Suchecki2005EPL,Suchecki2005PRE,Antal2006PRL,Sood2008PRE,Vazquez2008NewJPhys}, we first select a node $i$ to be updated with probability $1/N$. Then, a neighbor of $i$, denoted by $j$, is selected out of the neighbors of $i$ with the equal probability, i.e., the inverse of $i$'s degree. Then, $i$ copies $j$'s state.
Under the invasion process (IP) \cite{Castellano2005AIPConf,Antal2006PRL,Sood2008PRE}, we first select a parent node $i$ with probability $1/N$. Then, a neighbor of $i$, again denoted by $j$, is selected for updating with the probability equal to the inverse of $i$'s degree. Then, $j$ copies $i$'s state.
In the following,
we collectively refer to the dynamics under LD, VM, and IP as opinion dynamics.

The three update rules coincide for regular networks (i.e., networks in which all nodes have the same degree). In heterogeneous networks,
opinion dynamics including $\left<T\right>$ depends on the update rule
\cite{Castellano2005AIPConf,Antal2006PRL,Sood2008PRE}. In the present study,
we allow heterogeneous networks.

\section{Small networks maximizing the mean consensus time}\label{sec:exact}

\subsection{Methods}\label{sub:maximize methods}

In this section, we consider small networks and
exactly calculate $\left<T\right>$ for any given
network and numerically maximize $\left<T\right>$ by gradually morphing network structure
with $N$ (i.e., number of nodes) and $M$ (i.e., number of links)
fixed.

We refer to the collection of the states of the $N$ nodes as configuration. There are $2^N$ possible configurations because each node takes either state $\mathbf{0}$ or $\mathbf{1}$. Two configurations correspond to consensus.
The mean consensus time depends on the starting configuration. We denote a configuration by $[s_1 \; \cdots \; s_N]$, where $s_i\in \{0, 1\}$ is the state of node $i$ ($1\le i\le N$), and
the mean consensus time starting from configuration $[s_1 \; \cdots \; s_N]$ 
by $\left<T\right>_{[s_1\; \cdots\; s_N]}$. It should be noted that $\left<T\right>_{[0 \; \cdots \; 0 ]}=\left<T\right>_{[1\; \cdots \; 1]}=0$.

\begin{figure}
\begin{center}
\includegraphics[width=10cm]{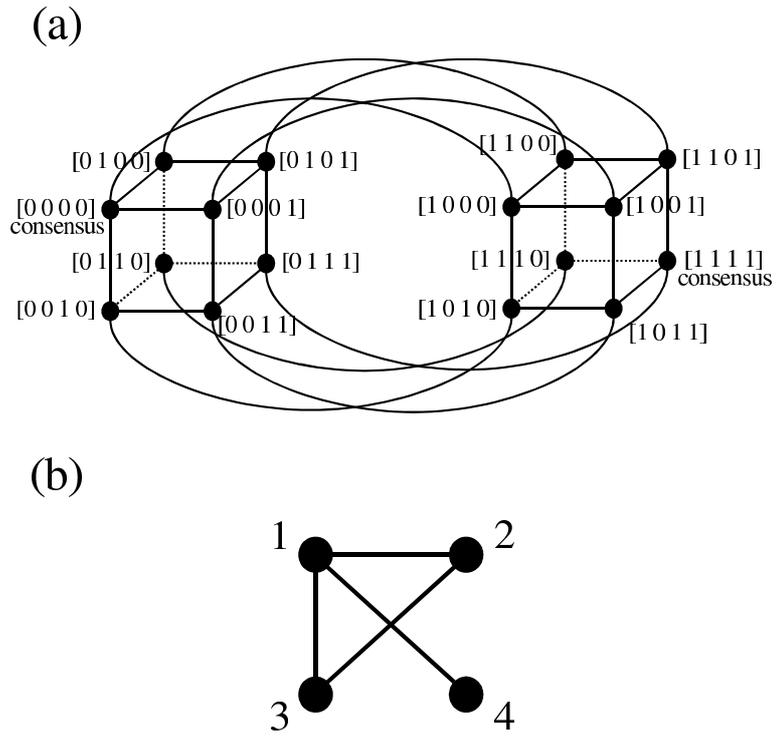}
\caption{(a) Schematic of transitions between configurations in opinion dynamics on a network possessing $N=4$ nodes. (b) A network with $N=4$.}
\label{fig:N=4}
\end{center}
\end{figure}

In a single update, the configuration may
switch to a neighboring configuration, where the neighborhood of a
configuration is defined as the configurations that differ from the
original configuration just in a single $s_i$. Each configuration has
$N$ neighbors, as shown in Fig.~\ref{fig:N=4}(a).

To explain the method for exactly calculating $\left<T\right>$,
we refer to the network with $N=4$ nodes shown in
Fig.~\ref{fig:N=4}(b)
and consider configuration [0101] (i.e., nodes 1 and 3 possess state $\mathbf{0}$, and nodes 2 and 4 possess state $\mathbf{1}$). Under LD, node 1 imitates node 2 in a single update with probability $1/8$, and node 1 imitates node 4 with probability $1/8$. If either of these events occurs, the configuration transits from [0101] to [1101]. If the link connecting nodes 1 and 3 is selected for the update, the configuration does not change. This event occurs with probability $1/4$. By taking into account all possible transitions from [0101] in a similar manner, we obtain
\begin{equation}
\left<T\right>_{[0101]} = \frac{1}{4}\left<T\right>_{[0101]} + \frac{1}{4}\left<T\right>_{[0001]} + \frac{1}{8}\left<T\right>_{[0100]} + \frac{1}{8}\left<T\right>_{[0111]} + \frac{1}{4}\left<T\right>_{[1101]} + \frac{1}{N}.
\label{eq:recursive T_[0101]}
\end{equation}
The last term on the right-hand side of Eq.~\eqref{eq:recursive T_[0101]}
results from the fact that an update consumes time $1/N$ by definition.

We can similarly derive the recursive equations for $2^N-2$ starting configurations in total, i.e., those except $[0000]$ and $[1111]$. By solving the set of the $2^N-2$ linear equations, we obtain the mean consensus time for all initial configurations. Finally, we define $\left<T\right>$ as the mean consensus time averaged over the initial configurations that possess the equal number of state $\mathbf{0}$ and $\mathbf{1}$ nodes. When $N$ is even, there are $N!/[(N/2)!]^2$ such configurations. When $N$ is odd, the average is taken over the $N!/\left\{[(N+1)/2]![(N-1)/2]!\right\}$ configurations in which the number of nodes in state $\mathbf{0}$ is smaller than that in state $\mathbf{1}$ just by one.

Using this exact method for calculating $\left<T\right>$, we numerically explored the network structure with the largest $\left<T\right>$ value for a given $N$, $M$, and update rule as follows.
\begin{enumerate}

\item Generate an initial network $G$ using one of the two following methods. In the first method, we prepare the star with $N$ nodes and $N-1$ links. Then, we add the remaining $M-N+1$ links between pairs of leaf (i.e., nonhub) nodes with the uniform density. In the second method, we prepare the chain with $N$ nodes and $N-1$ links. Then, we similarly add the remaining $M-N+1$ links between randomly selected nonadjacent pairs of nodes. In either method, we prohibit multiple links when adding the $M-N+1$ links. Networks generated by the two methods represent two extreme initial conditions.

\item Calculate $\left<T\right>$ for $G$ under the given update rule.

\item Select a link with the uniform probability $1/M$. Denote the endpoints of the selected link by $i$ and $j$. If the degrees of $i$ and $j$ are
at least two, delete the link, select a pair of nonadjacent nodes with the uniform probability, and create a link between the two nodes. If the degree of $i$ is equal to one and that of $j$ is at least two, we disconnect the link from $j$ but keep $i$ as an endpoint of the new link. We select the other endpoint of the new link with the uniform probability from the nodes that are not adjacent to
$i$ and connect it to $i$. It should be noted that either the degree of $i$ or that of $j$ is at least two because the network is in fact connected throughout the procedure. If the generated network is disconnected, we repeat the procedure until a connected network is generated. We refer to the new network as $G^{\prime}$.

\item Calculate $\left<T\right>$ for $G^{\prime}$.

\item If $\left<T\right>$ is larger for $G^{\prime}$ than $G$, we replace $G$ by $G^{\prime}$.

\item We repeat steps 3, 4, and 5 until the local maximum of $\left<T\right>$ is reached. In practice, if the rewiring does not occur in more than 5000 steps,
we stop repeating steps 3, 4, and 5,
calculate $\left<T\right>$, and record the network structure.

\end{enumerate}

We run simulations with five initial networks generated by each of the two methods (i.e., ten initial conditions in total). When the final network depends on the initial network, we select the one yielding the largest $\left<T\right>$ value. It should be noted that the obtained network realizes a local, but not global, maximum of $\left<T\right>$.

\subsection{Results}

First, we set $N=10$ and apply the optimization procedure described in Sec.~\ref{sub:maximize methods}.

\begin{figure}
\begin{center}
\includegraphics[width=12cm]{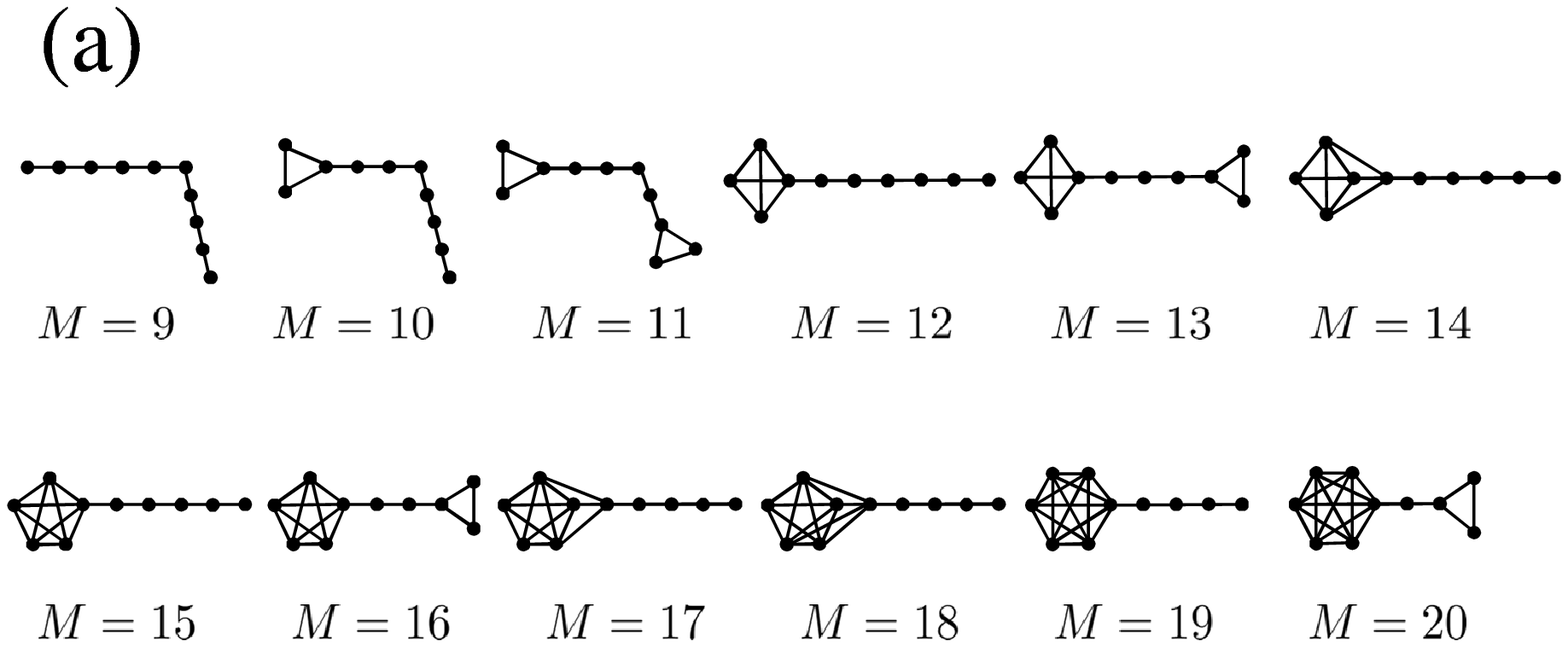}
\includegraphics[width=12cm]{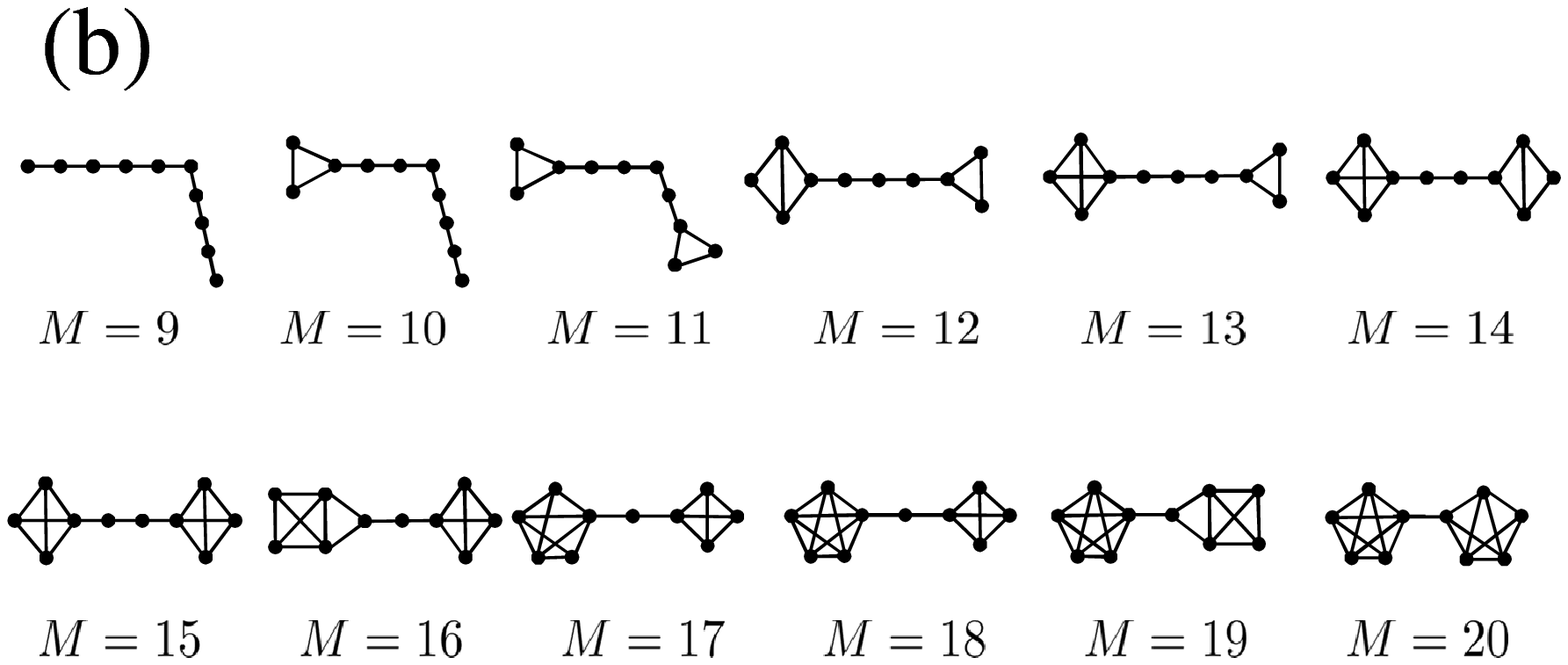}
\includegraphics[width=12cm]{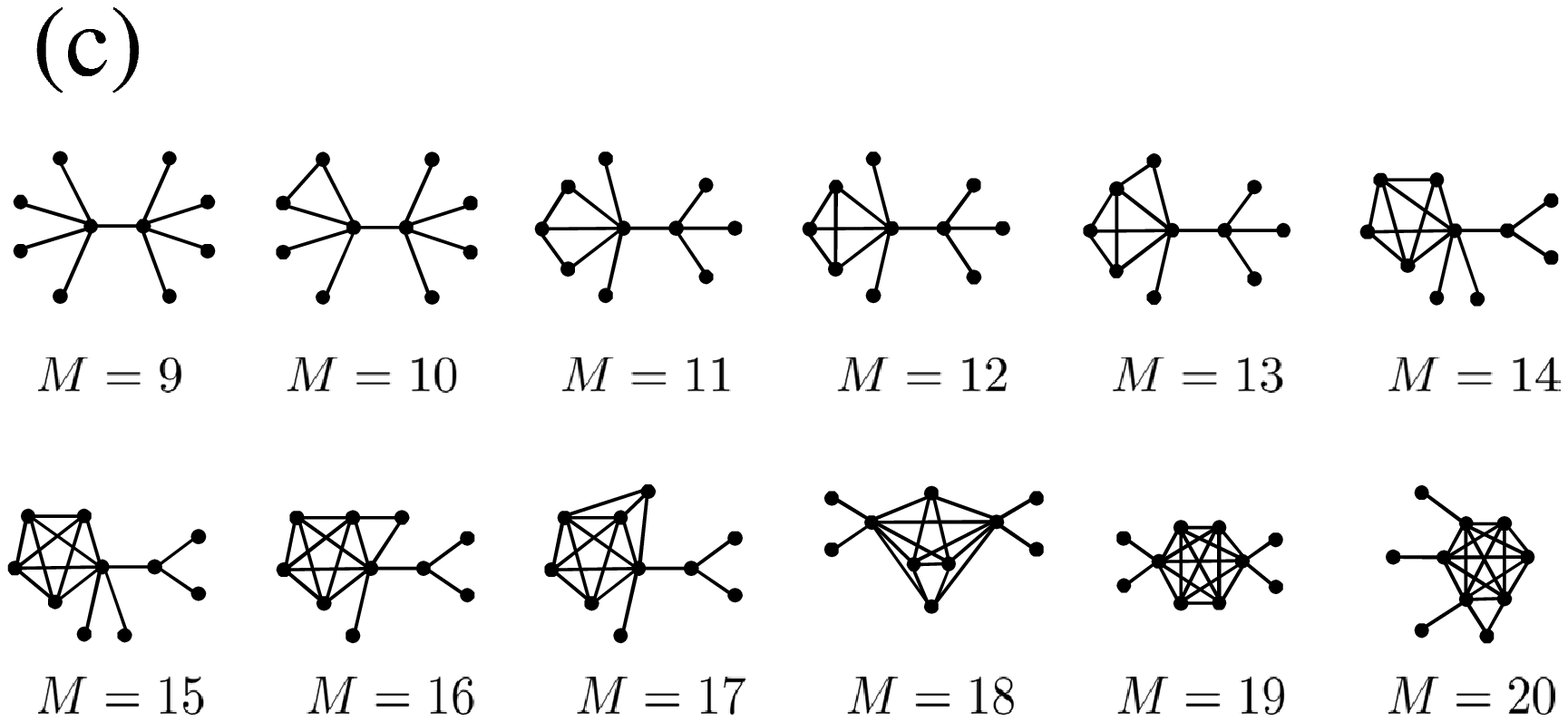}
\caption{Networks that maximize the mean consensus time for different numbers of links $M$. We set $N=10$. (a) LD, (b) VM, (c) IP.}
\label{fig:max T with N=10 shape}
\end{center}
\end{figure}

\begin{figure}
\begin{center}
\includegraphics[width=8cm]{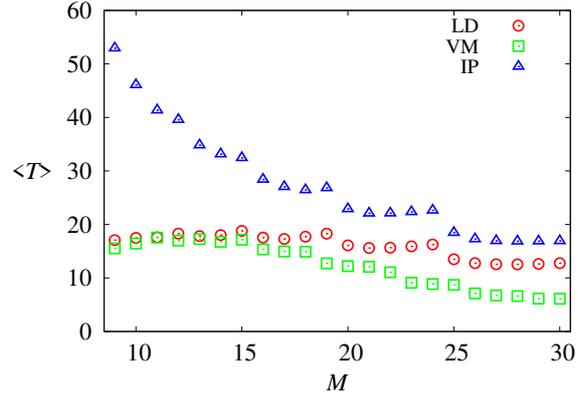}
\caption{Largest mean consensus time, $\left<T\right>$, with respect to network structure
when we vary $M (\ge N-1)$. We set $N=10$.}
\label{fig:max T with N=10}
\end{center}
\end{figure}

The networks maximizing $\left<T\right>$ under LD are shown for various $M$ values in Fig.~\ref{fig:max T with N=10 shape}(a).
The generated networks are close to the lollipop graph \cite{Lawler1986DiscMath,Chandra1989STOC,Brightwell1990RandStructAlgor}, which is defined as a network composed of a clique and a one-dimensional chain grafted to the clique. By definition, the lollipop graph can be only formed for specific values of $M$ given a value of $N$. Figure~\ref{fig:max T with N=10 shape}(a) shows that when $M$ are these values (i.e., $M=10$, 12, 15, 19), the network that maximizes $\left<T\right>$
is the lollipop graph. For other values of $M$, the obtained networks are close to the lollipop graph.
The maximized $\left<T\right>$ value is plotted against $M$ by the circles in
Fig.~\ref{fig:max T with N=10}. The peaks are located at the values of $M$ that enable the lollipop graph. Therefore, the lollipop graph is suggested to maximize $\left<T\right>$ under LD.

The networks maximizing $\left<T\right>$ under VM are shown in Fig.~\ref{fig:max T with N=10 shape}(b).
When $M=$ 11, 15, and 21, the generated networks are the barbell graph \cite{Mazo1982BellSysTechJ,Coppersmith1993SiamJDiscMath}, which is defined as the network possessing two cliques of the equal size connected by a chain. For the other $M$ values, $\left<T\right>$ is the largest for networks close to the barbell graph.
The maximized $\left<T\right>$ value is plotted against $M$ by the squares in Fig.~\ref{fig:max T with N=10}. The peaks of $\left<T\right>$ are located at $M=11$, 13, and 15. When $M=13$, the generated network is an asymmetric variant of the barbell graph in which one clique has three nodes and the other clique has four nodes [Fig.~\ref{fig:max T with N=10 shape}(b)]. Taken together with the results for $M=11$ and 15, the barbell graph is suggested to maximize $\left<T\right>$ under VM. 

\begin{figure}
\begin{center}
\includegraphics[width=6cm]{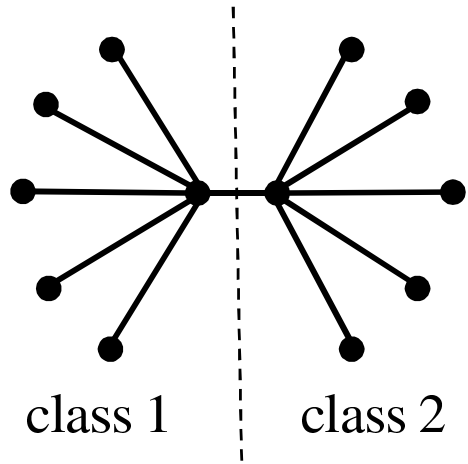}
\caption{Double-star graph.}
\label{fig:double-star}
\end{center}
\end{figure}

The networks maximizing $\left<T\right>$ under IP are shown in
Fig.~\ref{fig:max T with N=10 shape}(c). When $M=9$, which is the minimum possible value to keep the network connected, the generated network is the so-called double-star graph (Fig.~\ref{fig:double-star}), in which
two stars are connected by an additional link between the two hubs
\cite{Santos2008Nature}. When $M$ is slightly larger than $N-1$,
the generated networks are close to the double-star graph [Fig.~\ref{fig:max T with N=10 shape}(c)]. For larger $M$ values,
the generated networks are not similar to the double-star graph.
The maximized $\left<T\right>$ value is plotted against $M$ by the triangles in
Fig.~\ref{fig:max T with N=10}. The $\left<T\right>$ value monotonically decreases with $M$, suggesting that
the double-star graph maximizes $\left<T\right>$ under IP.

Next, for LD, we confine ourselves to the lollipop graph and search for
the lollipop graph with the largest value of $\left<T\right>$. To this end, we set
$N=15$ and $N=20$, and vary the size of the clique. We
are allowed to use larger values of $N$ as compared to the numerical simulations described in Figs.~\ref{fig:max T with N=10 shape} and \ref{fig:max T with N=10}, for which $N=10$, for two reasons. First, in the current set of numerical simulations, we do not have to maximize $\left<T\right>$ by gradually changing networks. Second, because of the symmetry inherent in the lollipop graph, we can considerably reduce the number of the linear equations to solve [e.g., 
Eq.~\eqref{eq:recursive T_[0101]}]. Concretely, we have to only maintain the number of nodes in the $\mathbf{1}$ state in the clique and the configuration of the chain, because all nodes in the clique are structurally equivalent. 

\begin{figure}
\begin{center}
\includegraphics[width=8cm]{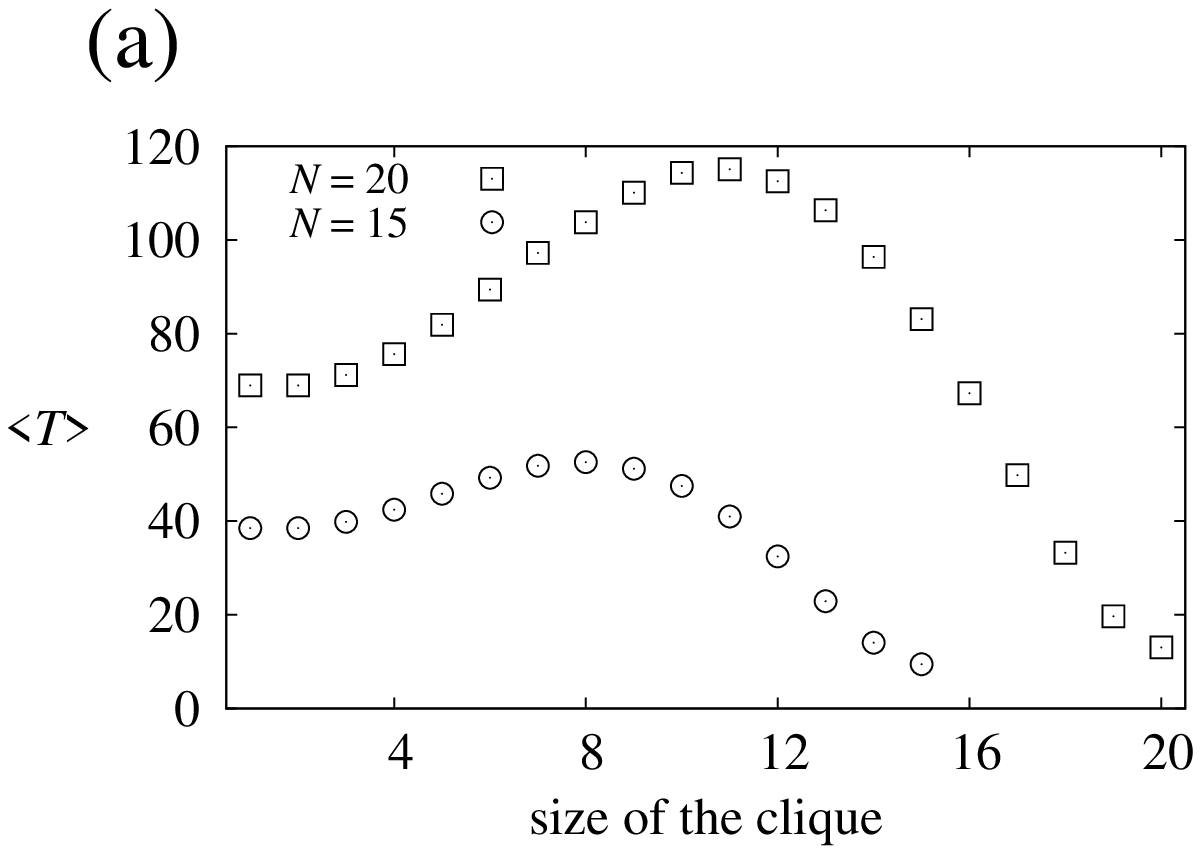}
\includegraphics[width=8cm]{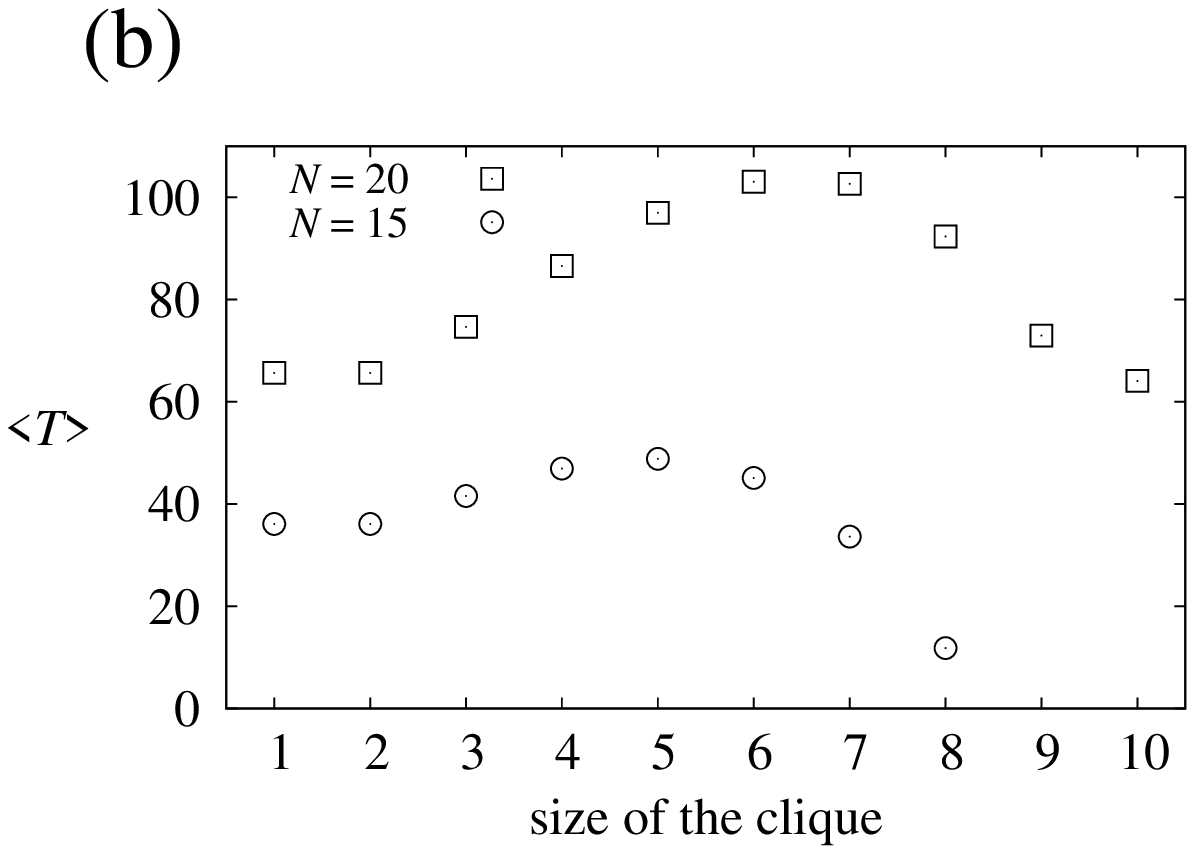}
\caption{(a) Mean consensus time, $\left<T\right>$, for the lollipop graph with various $M$ values under LD. (b) $\left<T\right>$ for the barbell graph with various $M$ values under VM. In (a) and (b), the horizontal axis represents the size of the clique in the lollipop or barbell graph. We set $N=15$ and $N=20$.}
\label{fig:max T with N=15 N=20}
\end{center}
\end{figure}

In Fig.~\ref{fig:max T with N=15 N=20}(a), $\left<T\right>$ is plotted as a function of the size of the clique in the lollipop graph when $N$ is fixed.
The figure indicates that $\left<T\right>$ is the largest for the lollipop graph having approximately half the nodes in the clique. The corresponding results for the barbell graph under VM are shown in Fig.~\ref{fig:max T with N=15 N=20}(b). The figure indicates that $\left<T\right>$ is the largest for the barbell graph that has approximately $N/3$ nodes in each clique and the chain.

\section{Analytical estimation of the mean consensus time}\label{sec:order}

In this section, we analytically assess the dependence of $\left<T\right>$ on $N$ for the lollipop, barbell, and double-star graphs under the three update rules. Our analysis is based on the probability and mean time for transitions among typical coarse-grained configurations.

\subsection{Lollipop graph}\label{sub:order lollipop}

Because we are interested in the asymptotic dependence of $\left<T\right>$ on $N$, we assume in this section that the lollipop graph contains $2N$ nodes; the clique and chain in the lollipop graph contain $N$ nodes each.
For all three update rules, consensus within the clique is reached in $O(N)$ time if the nodes in the chain do not affect the opinion dynamics within the clique.

In the following, we estimate the mean consensus time by assessing
approximate lower and upper bounds of $\left<T\right>$.
Typical configurations of the opinion dynamics on the lollipop graph are schematically shown in Fig.~\ref{fig:lollipop bounds}. An open and filled circle represents a node in the $\mathbf{0}$ and $\mathbf{1}$ states, respectively. The initial configuration is totally random and depicted as configuration I in 
Fig.~\ref{fig:lollipop bounds}.

\subsubsection{LD}\label{sub:lollipop LD}

\begin{figure}
\begin{center}
\includegraphics[width=10cm]{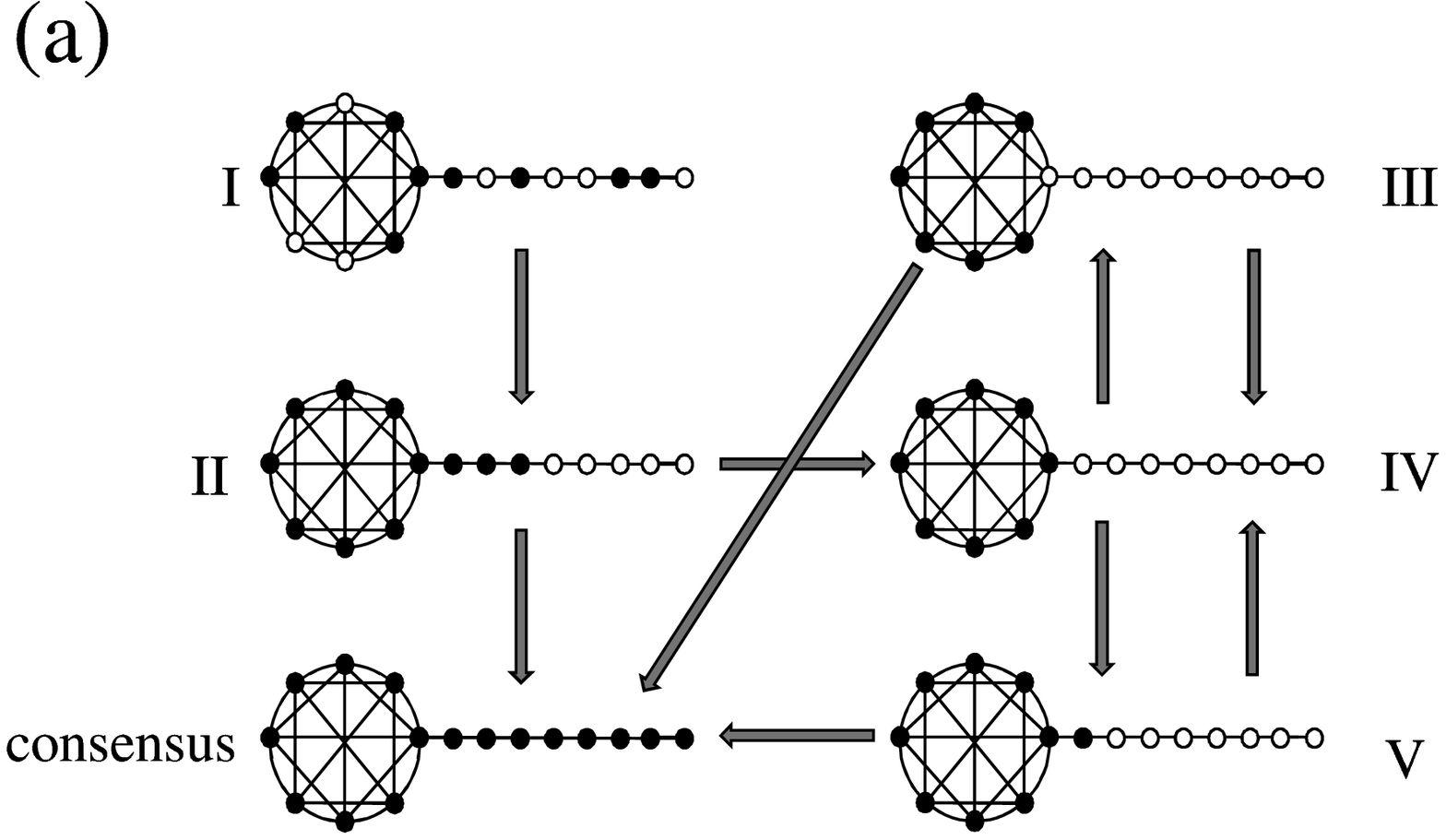}
\includegraphics[width=10cm]{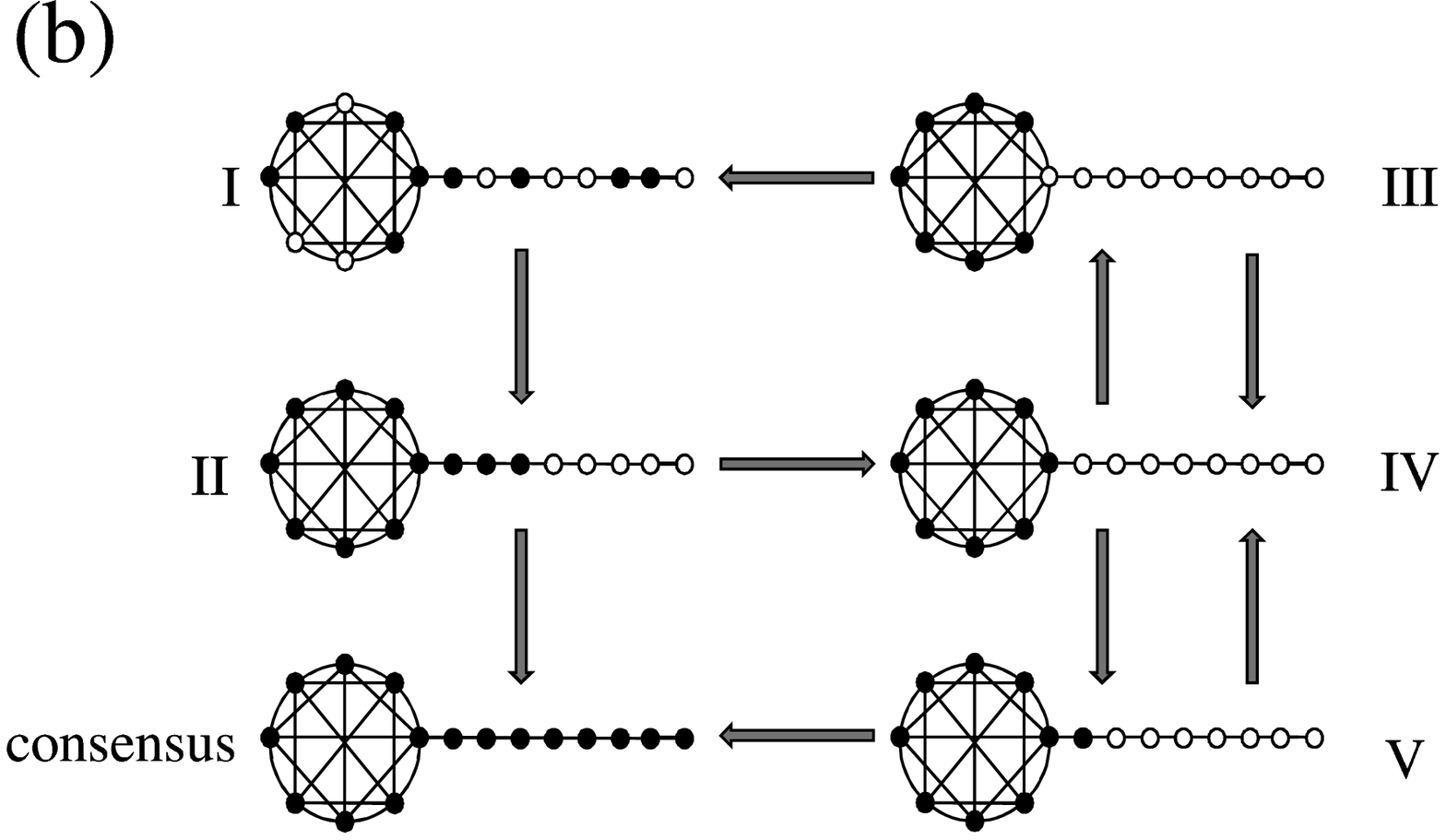}
\includegraphics[width=10cm]{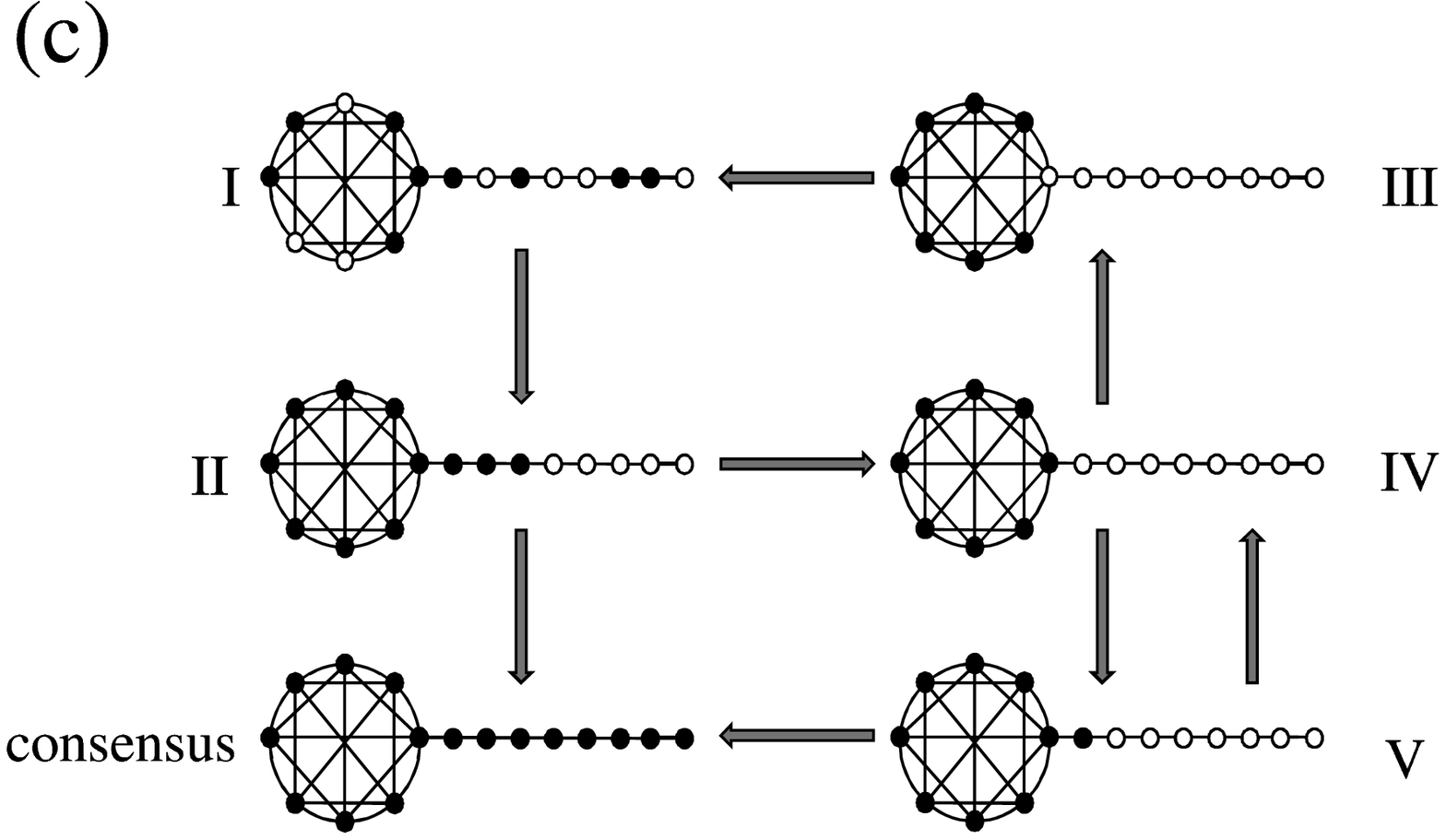}
\caption{Transitions among typical configurations
for evaluating approximate lower and upper bounds of $\left<T\right>$
for the lollipop graph.
(a) Schematic for evaluating a lower bound.
(b) Schematic for evaluating an upper bound under LD and IP.
(c) Schematic for evaluating an upper bound under VM.}
\label{fig:lollipop bounds}
\end{center}
\end{figure}

We call a consecutive segment on the chain occupied by the same state
(i.e., $\mathbf{0}$ or $\mathbf{1}$) the domain.
Under LD, a node in the chain has degree two (except for the one at the end of the chain, whose degree is one) and is updated once in $N(N-1)/2 + N=O(N^2)$ steps on average.
Because an update event consumes time $1/2N$ by definition, 
the mean time before a node on the chain is updated is $O(N)$. 
In the ordinary voter model on the one-dimensional chain, the domain size grows to an $O(N)$ size in $O(N^2)$ time \cite{Bramson1980AnnProb,CoxGriffeath1986AnnProb}.
Therefore, under LD, the time needed for domains to grow to an $O(N)$ size is equal to $O(N^3)$. By this time, consensus would be reached within the clique because consensus of the clique if it were isolated occurs in $O(N)$ time.
To summarize, the transition from configuration I to configuration II
occurs in $O(N^3)$ time.
It should be noted that configuration II shown in Fig.~\ref{fig:lollipop bounds}(a) includes configurations in which the chain
contains two or more domains of the opposite states with characteristic length $O(N)$. We obtain
\begin{equation}
\left<T\right>_{\rm I} = O(N^3) + \left<T\right>_{\rm II},
\label{eq:T_I lollipop LD}
\end{equation}
where $\left<T\right>_{\rm I}$ and
$\left<T\right>_{\rm II}$ are the mean consensus times starting from configurations I and II, respectively.
Likewise we define $\left<T\right>_{\rm III}$, $\left<T\right>_{\rm IV}$, and $\left<T\right>_{\rm V}$, where III, IV, and V correspond to the three configurations shown in Fig.~\ref{fig:lollipop bounds}(a).

There are two types of configurations that can be reached from configuration II. With a $O(1)$ probability $p^{\rm LD}$, which is in fact equal to the fraction of nodes on the chain that takes the same state as that of the clique \cite{Sood2008PRE}, configuration II transits to the consensus of the entire network [Fig.~\ref{fig:lollipop bounds}(a)]. This event requires $O(N^3)$ time because each node on the chain is updated once per $O(N)$ time and consensus of the chain requires $O(N^2)$ update events per node \cite{Cox1989AnnProb}.
Otherwise, starting from configuration II, the chain is eventually occupied by the state opposite to that taken by the clique [configuration IV shown in Fig.~\ref{fig:lollipop bounds}(a)] with probability $1-p^{\rm LD}=O(1)$. This event also requires $O(N^3)$ time. Therefore, we obtain
\begin{equation}
\left<T\right>_{\rm II} = p^{\rm LD} O(N^3) + \left( 1 - p^{\rm LD} \right) \left[ O(N^3) + \left<T\right>_{\rm IV} \right].
\label{eq:T_II lollipop LD}
\end{equation}

There are two possible types of events that occur in configuration IV. First, the state taken by the chain may invade a node in the clique (configuration III). Second, the state taken by the clique may invade the node on the chain adjacent to the clique (configuration V). Either event occurs with probability $1/2$ because the link connecting the clique and chain must be taken, and the direction of opinion transmission on the selected link
determines the type of the event. The occurrence of either event requires $O(N)$ time because the mentioned single link in the chain must be selected for either event to occur. Therefore, we obtain
\begin{equation}
\left<T\right>_{\rm IV} = \frac{1}{2} \left[ O(N) + \left<T\right>_{\rm III} \right] + \frac{1}{2} \left[ O(N) + \left<T\right>_{\rm V} \right].
\label{eq:T_IV lollipop LD}
\end{equation}

From configuration V, the consensus of the entire lollipop graph is reached in $O(N^3)$ time with probability $1/N$ because, under LD, the consensus probability of a state is equal to the number of nodes possessing the state in an arbitrary undirected network (chain in the present case) \cite{Antal2006PRL,Sood2008PRE}. With probability $(N-1)/N$, configuration IV is revisited. This event requires $O(N)$ updates per node, which corresponds to $O(N^2)$ time. These probabilities and times can be calculated as the hitting probability and time of the random walk on the chain with two absorbing boundaries \cite{Redner2001book,Vankampen2007book}. Therefore, we obtain
\begin{equation}
\left<T\right>_{\rm V} = \frac{1}{N} O(N^3) + \frac{N-1}{N} \left[ O(N^2) + \left<T\right>_{\rm IV} \right].
\label{eq:T_V lollipop LD}
\end{equation}

From configuration III, the single invader in the clique spreads its opinion in the clique to lead to consensus of the entire network after $O(N)$ time with probability approximately equal to $1/N$. This probability may be an overestimate because the clique and chain in fact interact possibly to prevent consensus from being reached in $O(N)$ time.
 Otherwise, configuration IV or a more complicated configuration as represented by configuration I is revisited. The transition to the latter 
configuration may occur because, starting from configuration III, the clique may experience a mixture of the ${\mathbf 0}$ and ${\mathbf 1}$ states for some time during which the clique transforms the chain to alternating small domains of the opposite 
states. However, to assess a lower bound of $\left<T\right>$, we pretend that
configuration IV is reached from configuration III with probability $(N-1)/N$ after short time $O(\ln N)$, which is true if the clique were disconnected from the chain \cite{BenAvraham1990JPhysA}. Then, we obtain
\begin{equation}
\left<T\right>_{\rm III} = \frac{1}{N} O(N) + \frac{N-1}{N} \left[ O(\ln N) + \left<T\right>_{\rm IV} \right].
\label{eq:T_III lollipop LD lower}
\end{equation}
Because consensus would require longer time if we start from configuration I than IV, we assume that Eq.~\eqref{eq:T_III lollipop LD lower} gives a lower bound of $\left<T\right>$.
By solving Eqs.~\eqref{eq:T_I lollipop LD}, \eqref{eq:T_II lollipop LD}, \eqref{eq:T_IV lollipop LD}, \eqref{eq:T_V lollipop LD}, and \eqref{eq:T_III lollipop LD lower},
we obtain $\left<T\right>_{\rm I}$, $\left<T\right>_{\rm II}$, 
$\left<T\right>_{\rm III}$, $\left<T\right>_{\rm IV}$, 
$\left<T\right>_{\rm V} = O(N^3)$ as a rough lower bound of $\left<T\right>$.
 
To estimate an upper bound of $\left<T\right>$, we consider the transitions between typical configurations shown in Fig.~\ref{fig:lollipop bounds}(b), which differs from Fig.~\ref{fig:lollipop bounds}(a) only in the transitions from configuration III. Now, with probability $q^{\rm LD}=1/N$, we presume that the configuration returns to a random one (configuration I), which is adversary to consensus. With probability $1-q^{\rm LD}$, we safely assume that configuration IV is revisited in time $O(\ln N)$. It should be noted that in time $O(\ln N)$, a link on the chain has rarely been selected for large $N$, such that the chain would not be divided into multiple domains of the opposite states in time $O(\ln N)$.
By combining Eqs.~\eqref{eq:T_I lollipop LD}, \eqref{eq:T_II lollipop LD},
\eqref{eq:T_IV lollipop LD}, \eqref{eq:T_V lollipop LD}, and
\begin{equation}
\left<T\right>_{\rm III} = q^{\rm LD}\left[\left<T\right>_{\rm I} + O(N)\right]
+ (1-q^{\rm LD})\left[\left<T\right>_{\rm IV}+O(\ln N)\right],
\label{eq:T_III lollipop LD upper}
\end{equation}
we obtain $\left<T\right>_{\rm I}$, $\left<T\right>_{\rm II}$,
$\left<T\right>_{\rm III}$, $\left<T\right>_{\rm IV}$, 
$\left<T\right>_{\rm V} = O(N^3)$ as a rough upper bound of $\left<T\right>$.
It should be noted that the same conclusion holds true as long as $q^{\rm LD}=O(1/N)$.

In fact, $\left<T\right>$ is exactly upper-bounded by $O(N^3)$ for arbitrary networks on the basis of the following alternative arguments. We denote the number of nodes in the $\mathbf{1}$ state by $N_1$. In an update event, $N_1$ increases or decreases by one with the same probability \cite{Antal2006PRL,Sood2008PRE}. With the remaining probability, $N_1$ does not change. Therefore, $N_1$ performs an unbiased random walk on interval $[0, N]$. The probability that $N_1$ changes in an update event is at least $O(1/N^2)$ because there are at most $N(N-1)/2$ links in the network. Because a single update event consumes time $1/N$ and the random walk hits either boundary (i.e., 0 or $N$) after $O(N^2)$ actual hops, we obtain $[(1/N)/O(1/N^2)]\times O(N^2) = O(N^3)$ as an upper bound of $\left<T\right>$.

On the basis of the rough lower and upper bounds, we conclude
$\left<T\right>=O(N^3)$ for the combination of the lollipop graph and LD.

\subsubsection{VM}

We evaluate approximate lower and upper bounds of $\left<T\right>$ under VM and IP in a similar manner. It should be noted that
a node in the chain is updated once per $O(1)$ time under VM and IP, which is in contrast to the case of LD.

To derive an approximate lower bound under VM, we consider the same transitions among typical configurations as those under LD [Fig.~\ref{fig:lollipop bounds}(a)].
We obtain
\begin{align}
&\left<T\right>_{\rm I} = O(N^2) + \left<T\right>_{\rm II},
\label{eq:T_I lollipop VM}\\
&\left<T\right>_{\rm II} = p^{\rm VM} O(N^2) + \left( 1 - p^{\rm VM} \right) \left[ O(N^2) + \left<T\right>_{\rm IV} \right],
\label{eq:T_II lollipop VM}\\
&\left<T\right>_{\rm III} = \frac{1}{N} O(N) + \frac{N-1}{N} \left[ O(\ln N) + \left<T\right>_{\rm IV} \right],
\label{eq:T_III lollipop VM lower}\\
&\left<T\right>_{\rm IV} = \frac{2}{N+2} \left[ O(N) + \left<T\right>_{\rm III} \right] + \frac{N}{N+2} \left[ O(1) + \left<T\right>_{\rm V} \right],
\label{eq:T_IV lollipop VM}\\
&\left<T\right>_{\rm V} = \frac{1}{2(N-1)} O(N^2) + \frac{2N-3}{2(N-1)} \left[ O(N) + \left<T\right>_{\rm IV} \right],
\label{eq:T_V lollipop VM}
\end{align}
where $p^{\rm VM}$, $1-p^{\rm VM} = O(1)$. To derive Eq.~\eqref{eq:T_IV lollipop VM}, we used the fact that, in a single update event, the transition from configuration IV to III occurs with probability $(1/2N)\times (1/N)$ and that from configuration IV to V occurs with probability $(1/2N)\times (1/2)$. To derive Eq.~\eqref{eq:T_V lollipop VM}, we used the fact that the fixation probability for a node $i$ (i.e., probability that the consensus of the state taken by $i$ is reached when all the other $N-1$ nodes possess the opposite state) under VM is proportional to $i$'s degree \cite{Antal2006PRL,Sood2008PRE}. Consider an isolated chain of length $N$ in which the leftmost node is in state $\mathbf{1}$ and all the other $N-1$ nodes are in state $\mathbf{0}$, as schematically
shown in configuration V in Fig.~\ref{fig:lollipop bounds}(a). Then, $\mathbf{0}$ and $\mathbf{1}$ fixate with probability $(2N-3)/[2(N-1)]$ and $1/[2(N-1)]$, respectively; $2(N-1)$ is equal to the sum of the degree of all nodes.
Equations~\eqref{eq:T_I lollipop VM}, \eqref{eq:T_II lollipop VM}, \eqref{eq:T_III lollipop VM lower}, \eqref{eq:T_IV lollipop VM}, and \eqref{eq:T_V lollipop VM} yield $\left<T\right>_{\rm I}$, $\left<T\right>_{\rm II}$, $\left<T\right>_{\rm III}$, $\left<T\right>_{\rm IV}$, $\left<T\right>_{\rm V} = O(N^2)$.

To estimate an upper bound of $\left<T\right>$, we consider transitions among typical configurations shown in Fig.~\ref{fig:lollipop bounds}(c). It is different from the diagram for LD [Fig.~\ref{fig:lollipop bounds}(b)] in that the transition from configuration III to configuration IV is not present.
We modified the diagram because, under VM, the node that belongs to the chain and is adjacent to the clique imitates the state of a node in the 
clique every $O(1)$ time. Therefore, even within time $O(\ln N)$, nodes in the chain may flip the state many times such that configuration IV is rarely
visited directly from configuration III.
Given this situation, we replace Eq.~\eqref{eq:T_III lollipop VM lower} by
\begin{equation}
\left<T\right>_{\rm III} = O(N) + \left<T\right>_{\rm I}.
\label{eq:T_III lollipop VM upper}
\end{equation}
In other words, once configuration III is reached, we assume that configuration I is always reached after $O(N)$ time. Because
configuration I is considered to be adversary to consensus as compared to configuration IV, we assume that Eq.~\eqref{eq:T_III lollipop VM upper} gives an upper bound of $\left<T\right>$. The $O(N)$ time on the right-hand side of Eq.~\eqref{eq:T_III lollipop VM upper} comes from the fact that the opinion dynamics in an isolated clique of size $N$ ends up with consensus in $O(N)$ time.
Using Eqs.~\eqref{eq:T_I lollipop VM}, \eqref{eq:T_II lollipop VM}, \eqref{eq:T_IV lollipop VM}, \eqref{eq:T_V lollipop VM}, and \eqref{eq:T_III lollipop VM upper}, we obtain
$\left<T\right>_{\rm I}$, $\left<T\right>_{\rm II}$, $\left<T\right>_{\rm III}$, $\left<T\right>_{\rm IV}$, $\left<T\right>_{\rm V} = O(N^2)$.

Therefore, we conclude $\left<T\right>=O(N^2)$ for the combination of the lollipop graph and VM.

\subsubsection{IP}

For an approximate lower bound under IP, we obtain
\begin{align}
&\left<T\right>_{\rm I} = O(N^2) + \left<T\right>_{\rm II},
\label{eq:T_I lollipop IP}\\
&\left<T\right>_{\rm II} = p^{\rm IP} O(N^2) + \left( 1 - p^{\rm IP} \right) \left[ O(N^2) + \left<T\right>_{\rm IV} \right],
\label{eq:T_II lollipop IP}\\
&\left<T\right>_{\rm III} = \frac{1}{N} O(N) + \frac{N-1}{N} \left[ O(\ln N) + \left<T\right>_{\rm IV} \right],
\label{eq:T_III lollipop IP lower}\\
&\left<T\right>_{\rm IV} = \frac{N}{N+2} \left[ O(1) + \left<T\right>_{\rm III} \right] + \frac{2}{N+2} \left[ O(N) + \left<T\right>_{\rm V} \right],
\label{eq:T_IV lollipop IP}\\
&\left<T\right>_{\rm V} = \frac{2}{N+2} O(N^2) + \frac{N}{N+2} \left[ O(N) + \left<T\right>_{\rm IV} \right],
\label{eq:T_V lollipop IP}
\end{align}
where $p^{\rm IP}$, $1-p^{\rm IP} = O(1)$. To derive Eq.~\eqref{eq:T_IV lollipop IP}, we used the fact that, in a single update event, configuration IV transits to III with probability $(1/2N)\times (1/2)$ and V with probability $(1/2N)\times (1/N)$. To derive Eq.~\eqref{eq:T_V lollipop IP}, we used the fact that the fixation probability under IP is inversely proportional to the degree of the node \cite{Antal2006PRL,Sood2008PRE}. Similar to the case of VM, consider an isolated chain of length $N$ in which the leftmost node is in state $\mathbf{1}$ and all the other $N-1$ nodes are in state $\mathbf{0}$. Then, $\mathbf{0}$ and $\mathbf{1}$ fixate with probability $\propto (N-2)\times (1/2) + 1\times (1/1)$ and $\propto 1\times (1/1)$, respectively.
Equations~\eqref{eq:T_I lollipop IP}, \eqref{eq:T_II lollipop IP},
\eqref{eq:T_III lollipop IP lower}, \eqref{eq:T_IV lollipop IP}, and
\eqref{eq:T_V lollipop IP} yield
$\left<T\right>_{\rm I}$, $\left<T\right>_{\rm II}=O(N^2)$ and
$\left<T\right>_{\rm III}$, $\left<T\right>_{\rm IV}$, $\left<T\right>_{\rm V}
=  O(N\ln N)$. Because a random initial condition yields
$\left<T\right>_{\rm I}=O(N^2)$, we regard $O(N^2)$
as a rough lower bound of $\left<T\right>$.

To estimate an upper bound of $\left<T\right>$, we consider the same transitions among typical configurations as those for LD [Fig.~\ref{fig:lollipop bounds}(b)] and replace Eq.~\eqref{eq:T_III lollipop IP lower} by
\begin{equation}
\left<T\right>_{\rm III} = q^{\rm IP}\left[\left<T\right>_{\rm I}+O(N)\right]
+ (1-q^{\rm IP})\left[\left<T\right>_{\rm IV}+O(\ln N)\right],
\label{eq:T_III lollipop IP upper}
\end{equation}
where $q^{\rm IP}=1/N$.
To derive Eq.~\eqref{eq:T_III lollipop IP upper}, we assumed that configuration IV is reached in $O(\ln N)$ time with probability $1-q^{\rm IP}=(N-1)/N$. Within this time, the state monopolizing the clique would not invade the chain because such an event requires $O(N)$ time to occur.
By combining Eqs.~\eqref{eq:T_I lollipop IP}, \eqref{eq:T_II lollipop IP}, \eqref{eq:T_IV lollipop IP}, \eqref{eq:T_V lollipop IP}, and \eqref{eq:T_III lollipop IP upper}, we obtain
$\left<T\right>_{\rm I}$, $\left<T\right>_{\rm II}$, $\left<T\right>_{\rm III}$, $\left<T\right>_{\rm IV}$, $\left<T\right>_{\rm V} = O(N^2)$ as a rough upper bound of $\left<T\right>$.

Therefore, we conclude $\left<T\right>=O(N^2)$ for the combination of the lollipop graph and IP.

\subsection{Barbell graph}\label{sub:order barbell}

In the barbell graph, the two cliques may first reach the unanimity of the opposite states. This phenomenon may delay consensus, which is the case for the two-clique graph \cite{Sood2008PRE,Masuda2014arxiv_2cliques}. For simplicity, we assume in this section that each clique and the chain have $N$ nodes each such that the barbell graph is composed of $3N$ nodes.
We evaluate approximate lower and upper bounds for each update rule in the manner similar to that for the lollipop graph. Typical configurations and the transitions among them for the barbell graph are schematically shown in
Fig.~\ref{fig:barbell bounds}.

\subsubsection{LD}

\begin{figure}
\begin{center}
\includegraphics[width=10cm]{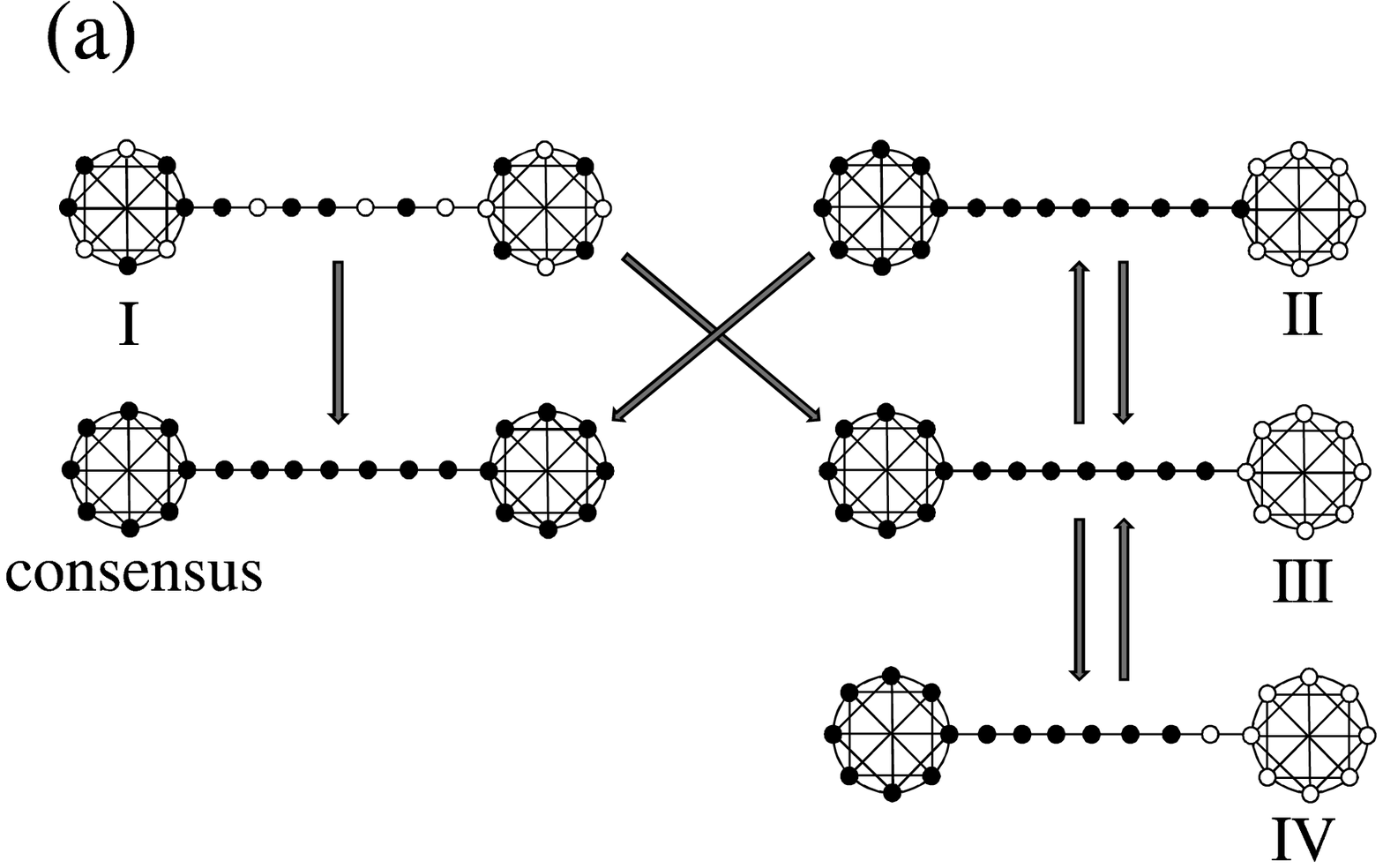}
\includegraphics[width=10cm]{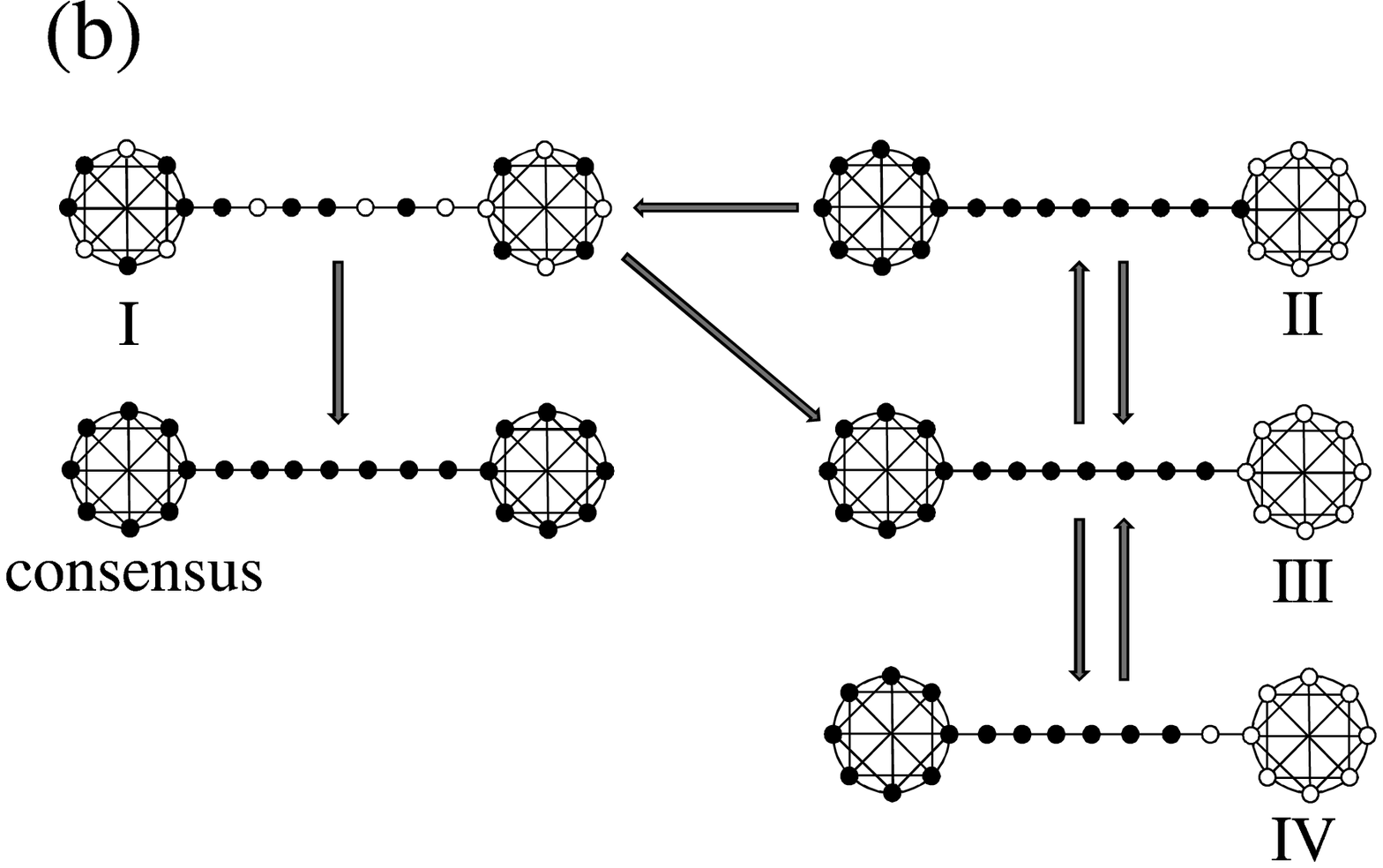}
\caption{Transitions among typical configurations
for evaluating approximate lower and upper bounds of $\left<T\right>$
for the barbell graph.
(a) Schematic for evaluating a lower bound.
(b) Schematic for evaluating an upper bound under LD and IP.
(c) Schematic for evaluating an upper bound under VM.}
\label{fig:barbell bounds}
\end{center}
\end{figure}

Under LD, a node belonging to the chain is updated once per $O(N)$ time. Therefore, the consensus of the chain would require $O(N^3)$ time if it were isolated. When the consensus of the chain is reached, each clique would have also reached consensus because the consensus within a clique occurs in $O(N)$ time. When two cliques end up with the same state after $O(N^3)$ time, the consensus of the entire barbell graph is realized. Otherwise, configuration III shown in Fig.~\ref{fig:barbell bounds}(a) is reached. Either event occurs with probability $1/2$. Therefore, we obtain
\begin{equation}
\left<T\right>_{\rm I} = \frac{1}{2} O(N^3) + \frac{1}{2} \left[ O(N^3) + \left<T\right>_{\rm III} \right].
\label{eq:T_I barbell LD}
\end{equation}

Starting from configuration III, the next change in the state occurs at the boundary between the chain and one of the two cliques whose state is opposite to that of the chain. Similar to the transition from configuration IV in the case of the lollipop graph (Sec.~\ref{sub:lollipop LD}), we obtain
\begin{equation}
\left<T\right>_{\rm III} = \frac{1}{2} \left[ O(N) + \left<T\right>_{\rm II} \right]+ \frac{1}{2} \left[ O(N) + \left<T\right>_{\rm IV} \right].
\label{eq:T_III barbell LD}
\end{equation}

Starting from configuration IV [Fig.~\ref{fig:barbell bounds}(a)], the state taken by just one node in the chain, which is adjacent to a clique, fixates in the chain in $O(N^3)$ time with probability $1/N$. Otherwise, with probability $(N-1)/N$, the state taken by $N-1$ nodes in the chain fixates in the chain in $O(N^2)$ time. In both cases, configuration III is recovered because of the symmetry. Therefore, we obtain
\begin{equation}
\left<T\right>_{\rm IV} = O(N^2) + \left<T\right>_{\rm III}.
\label{eq:T_IV barbell LD}
\end{equation}

We proceed similarly to the case of the lollipop graph to
evaluate an approximate lower bound of $\left<T\right>$. In other words,
as shown in Fig.~\ref{fig:barbell bounds}(a),
starting from configuration II, we suppose that
the consensus of the entire network is attained with probability $1/N$ in $O(N)$ time and that configuration III is revisited in $O(\ln N)$ time with the remaining probability. Then, we obtain
\begin{equation}
\left<T\right>_{\rm II} = \frac{1}{N} O(N) + \frac{N-1}{N} \left[ O(\ln N) + \left<T\right>_{\rm III} \right].
\label{eq:T_II barbell LD}
\end{equation}
By combining Eqs.~\eqref{eq:T_I barbell LD}, \eqref{eq:T_III barbell LD},
\eqref{eq:T_IV barbell LD}, and \eqref{eq:T_II barbell LD},
we obtain $\left<T\right>_{\rm I}, \left<T\right>_{\rm II}, \left<T\right>_{\rm III}, \left<T\right>_{\rm IV} = O(N^3)$.

To evaluate an approximate upper bound, we only modify the transitions from configuration II, similar to the case of the lollipop graph. 
Without loss of generality, we assume that the chain and the first clique is occupied by the $\mathbf{1}$ state and that the second clique has a single node in state $\mathbf{1}$ and $N-1$ nodes in state $\mathbf{0}$, as illustrated
by configuration II in Fig.~\ref{fig:barbell bounds}(a).
With a small probability $r^{\rm LD}=O(1/N)$, the $\mathbf{1}$ state proliferates in the second clique to possibly occupy it in $O(N)$ time.
Because a node in the chain is updated once per $O(N)$ time, 
the configuration of the chain may turn into a mixture of states $\mathbf{0}$ and $\mathbf{1}$ in the $O(N)$ time. Therefore, with probability $r^{\rm LD}$, we assume that configuration II transits to configuration I.
With probability $1-r^{\rm LD}$, the single node in state $\mathbf{1}$ is extinguished in the second clique in $O(\ln N)$ time such that configuration III is revisited. It should be noted that a node in the chain is rarely updated in $O(\ln N)$ time such that the unanimity on the chain is not perturbed in the $O(\ln N)$ time. By collecting these contributions, we obtain
\begin{equation}
\left<T\right>_{\rm II} = r^{\rm LD} \left[ O(N) + \left<T\right>_{\rm I} \right] + \left(1-r^{\rm LD}\right) \left[ O(\ln N) + \left<T\right>_{\rm III} \right].
\label{eq:T_II barbell LD upper}
\end{equation}
By combining Eqs.~\eqref{eq:T_I barbell LD}, \eqref{eq:T_III barbell LD}, \eqref{eq:T_IV barbell LD}, and \eqref{eq:T_II barbell LD upper}, we obtain
$\left<T\right>_{\rm I}, \left<T\right>_{\rm II}, \left<T\right>_{\rm III}, \left<T\right>_{\rm IV} = O(N^3)$.
It should be noted that the arguments based on the unbiased random walk (Sec.~\ref{sub:lollipop LD}) also lead to same upper bound of $\left<T\right>$.

Therefore, we conclude
$\left<T\right>=O(N^3)$ for the combination of the barbell graph and LD.

\subsubsection{VM}

For an approximate lower bound of $\left<T\right>$ under VM, we assume the same types of transitions among configurations as those for LD [Fig.~\ref{fig:barbell bounds}(a)] to obtain
\begin{align}
&\left<T\right>_{\rm I} = \frac{1}{2} O(N^2) + \frac{1}{2} \left[ O(N^2) + \left<T\right>_{\rm III} \right],
\label{eq:T_I barbell VM}\\
&\left<T\right>_{\rm II} = \frac{1}{N} O(N) + \frac{N-1}{N} \left[ O(\ln N) + \left<T\right>_{\rm III} \right],
\label{eq:T_II barbell VM lower}\\
&\left<T\right>_{\rm III} = \frac{2}{N+2} \left[ O(N) + \left<T\right>_{\rm II} \right]+ \frac{N}{N+2} \left[ O(1) + \left<T\right>_{\rm IV} \right],
\label{eq:T_III barbell VM}\\
&\left<T\right>_{\rm IV} = O(N) + \left<T\right>_{\rm III},
\label{eq:T_IV barbell VM}
\end{align}
leading to $\left<T\right>_{\rm I}, \left<T\right>_{\rm II}, \left<T\right>_{\rm III}, \left<T\right>_{\rm IV} = O(N^3)$.

To estimate an upper bound, we assume the transitions among configurations shown in Fig.~\ref{fig:barbell bounds}(c). It is different from those for LD [Fig.~\ref{fig:barbell bounds}(b)] in the transitions starting from configuration II. 
Similar to the case of the LD, we assume without loss of generality that
the chain, first clique, and a single node in the second clique are in state $\mathbf{1}$ and the other $N-1$ nodes in the second clique are in state $\mathbf{0}$. The $\mathbf{1}$ state in the second clique fixates there with probability
$r^{\rm VM}=O(1/N)$ in $O(N)$ time and 
is eradicated with probability $1-r^{\rm VM}$ in $O(\ln N)$ time.
Because each node in the chain is updated once per unit time, domains of characteristic length $O(\sqrt{\ln N})$ are formed within the $O(\ln N)$ time in the latter situation. The resulting configuration is shown as configuration V in Fig.~\ref{fig:barbell bounds}(c). Given configuration V, the consensus of the chain occurs in $O(N^2)$ time, such that configuration III is revisited.
By combining Eqs.~\eqref{eq:T_I barbell VM}, \eqref{eq:T_III barbell VM}, \eqref{eq:T_IV barbell VM}, and
\begin{equation}
\left<T\right>_{\rm II} = r^{\rm VM} \left[ O(N) + \left<T\right>_{\rm I} \right] + (1-r^{\rm VM}) \left[ O(\ln N) + O(N^2) + \left<T\right>_{\rm III} \right],
\label{eq:T_II barbell VM upper}
\end{equation}
we obtain $\left<T\right>_{\rm I}, \left<T\right>_{\rm II}, \left<T\right>_{\rm III}, \left<T\right>_{\rm IV} = O(N^3)$.

Therefore, we conclude
$\left<T\right>=O(N^3)$ for the combination of the barbell graph and VM.

\subsubsection{IP}

For an approximate lower bound of $\left<T\right>$ under IP, we consider Fig.~\ref{fig:barbell bounds}(a) to obtain
\begin{align}
&\left<T\right>_{\rm I} = \frac{1}{2} O(N^2) + \frac{1}{2} \left[ O(N^2) + \left<T\right>_{\rm III} \right],
\label{eq:T_I barbell IP}\\
&\left<T\right>_{\rm II} = \frac{1}{N} O(N) + \frac{N-1}{N} \left[ O(\ln N) + \left<T\right>_{\rm III} \right],
\label{eq:T_II barbell IP lower}\\
&\left<T\right>_{\rm III} = \frac{N}{N+2} \left[ O(1) + \left<T\right>_{\rm II} \right]+ \frac{2}{N+2} \left[ O(N) + \left<T\right>_{\rm IV} \right],
\label{eq:T_III barbell IP}\\
&\left<T\right>_{\rm IV} = O(N) + \left<T\right>_{\rm III},
\label{eq:T_IV barbell IP}
\end{align}
leading to $\left<T\right>_{\rm I} = O(N^2)$
and $\left<T\right>_{\rm II}, \left<T\right>_{\rm III}, \left<T\right>_{\rm IV} = O(N\ln N)$. Because a random initial condition yields
$\left<T\right>_{\rm I}=O(N^2)$, we regard $O(N^2)$
as a rough lower bound of $\left<T\right>$.

For an approximate upper bound, we consider the same diagram as that
for LD [Fig.~\ref{fig:barbell bounds}(b)]. The rationale behind this
choice is that the unanimity in the chain in configuration II is not
disturbed in $O(\ln N)$ time. This holds true because it takes $O(N)$
time before the state taken by a clique may invade a node that
belongs to the chain and is adjacent to the clique.  Therefore, we
replace Eq.~\eqref{eq:T_II barbell IP lower} by
\begin{equation}
\left<T\right>_{\rm II} = r^{\rm IP} \left[ O(N) + \left<T\right>_{\rm I} \right] + \left(1-r^{\rm IP}\right) \left[ O(\ln N) + \left<T\right>_{\rm III} \right],
\label{eq:T_II barbell IP upper}
\end{equation}
where $r^{\rm IP}=O(1/N)$.
By combining Eqs.~\eqref{eq:T_I barbell IP}, \eqref{eq:T_III barbell IP},
\eqref{eq:T_IV barbell IP}, and \eqref{eq:T_II barbell IP upper},
we obtain $\left<T\right>_{\rm I}, \left<T\right>_{\rm II}, \left<T\right>_{\rm III}, \left<T\right>_{\rm IV} = O(N^2)$.

Therefore, we conclude
$\left<T\right>=O(N^2)$ for the combination of the barbell graph and IP.

\subsection{Double-star graph}\label{sub:order double-star}

We evaluate the mean consensus time for the double-star graph under the three update rules using a different method from that for the lollipop and barbell graphs. It is mathematically established that the so-called dual process of the opinion dynamics is the so-called coalescing random walk \cite{Liggett1985book,Donnelly1983MPCPS,Durrett1988book,Cox1989AnnProb}, which is defined as follows. Consider $N$ simple random walkers, with one walker located at each node initially. Walkers that have arrived at the same node are assumed to coalesce into one. Then, all $N$ walkers eventually coalesce into one in a finite network. The dependence on the update rule only appears in the rule with which we move the walkers \cite{Donnelly1983MPCPS,MasudaOhtsuki2009NewJPhys}. The mathematical duality between the opinion dynamics and coalescing random walk guarantees that the time at which the last two walkers coalesce is equal to the consensus time of the opinion dynamics \cite{Liggett1985book,Donnelly1983MPCPS,Durrett1988book,Cox1989AnnProb}.
The time to the coalescence of the last two walkers is considered to dominate the entire coalescing random walk process starting from the $N$ walkers and ending when the last two walkers have coalesced. Therefore, in this section, we assess $\left<T\right>$ by measuring the mean time at which two walkers starting from different nodes in the double-star graph meet.
We used the same technique for a different network in a previous study \cite{Masuda2014arxiv_2cliques}.

Consider the double-star graph with $2N$ nodes as shown in
Fig.~\ref{fig:double-star}. We call two symmetric parts composed of $N$ nodes the classes 1 and 2. Each class contains one hub node with degree $N$ and $N-1$ leaf nodes with degree 1.

\begin{figure}
\begin{center}
\includegraphics[width=8cm]{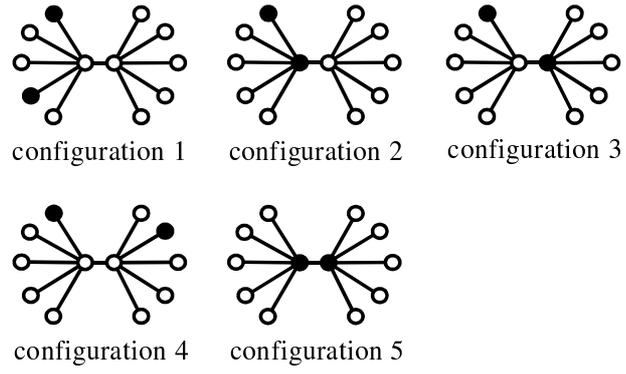}
\caption{Five configurations in the double-star graph that define $p_1(t)$, $\ldots$, $p_5(t)$.}
\label{fig:5 configs double-star}
\end{center}
\end{figure}

We define 
\begin{equation}
\bm{p}(t) = \begin{pmatrix}
      p_1(t)\\ p_2(t)\\ p_3(t)\\ p_4(t)\\ p_5(t)
\end{pmatrix},
\label{eq:def p(t)}
\end{equation}
where $t$ denotes the time. In Eq.~\eqref{eq:def p(t)}, $p_1(t)$ is the probability that the two walkers are located at different leaves in a single class (i.e., class 1 or 2) at time $t$ (configuration 1 shown in Fig.~\ref{fig:5 configs double-star}). $p_2(t)$ is the probability that
one walker stays in a class 1 leaf and the other walker stays in the class 1 hub, or one walker stays in a class 2 leaf and the other walker stays in the class 2 hub (configuration 2). $p_3(t)$ is the probability that one walker stays in a class 1 leaf and the other walker stays in the class 2 hub, or one walker stays in a class 2 leaf and the other walker stays in the class 1 hub (configuration 3).
$p_4(t)$ is the probability that a walker stays in a class 1 leaf and the other walker stays in a class 2 leaf (configuration 4). Finally,
$p_5(t)$ is the probability that a walker stays in the class 1 hub and the other walker stays in the class 2 hub (configuration 5).
We denote by $q(t)$ the probability that the two walkers meet at time $t$.

\subsubsection{LD}\label{sub:double LD}

Under LD, a link with one of the two directions is selected
with probability $1/[2(2N-1)]$ in an update event, which consumes time $1/2N$.
Therefore, we obtain
\begin{align}
  &\bm{p}\left(t+\frac{1}{2N}\right) = A^{{\rm LD}} \bm{p}(t), \\
  &q\left(t+\frac{1}{2N}\right) = \frac{1}{2(2N-1)} 
    \begin{pmatrix}
      0 & 2 & 0 & 0 & 2
    \end{pmatrix}
   \bm{p}(t),
\end{align}
where
\begin{equation}
	A^{{\rm LD}} = \frac{1}{2(2N-1)} 
    \begin{pmatrix}
      4(N-1) & N-2 & 0 & 0 & 0 \\
      2 & 3(N-1) & 1 & 0 & 0 \\
      0 & 1 & 3(N-1) & 2 & 2(N-1) \\
      0 & 0 & N-1 & 4(N-1) & 0 \\
      0 & 0 & 1 & 0 & 2(N-1)
    \end{pmatrix}. \label{ALD}
\end{equation}
Equation~\eqref{ALD} leads to
\begin{align}
	&(I - A^{{\rm LD}})^{-1} = \frac{2N-1}{2N+3} 
    \begin{pmatrix}
      N^2 + N + 1 & (N+1)(N-2) & N(N-2) & N(N-2) & (N-1)(N-2) \\
      2(N+1) & 2(N+1) & 2N & 2N & 2(N-1) \\
      2N & 2N & 6N & 6N & 6(N-1) \\
      N(N-1) & N(N-1) & 3N(N-1) & 3N^2-N+3 & 3(N-1)^2 \\
      1 & 1 & 3 & 3 & 5
    \end{pmatrix} \label{IALD}
\end{align}
and
\begin{equation}
	\frac{1}{2(2N-1)}
    \begin{pmatrix}
      0 & 2 & 0 & 0 & 2
    \end{pmatrix} (I - A^{{\rm LD}})^{-1} \\
  = \begin{pmatrix}
      1 & 1 & 1 & 1 & 1
    \end{pmatrix}. \label{11111LD}
\end{equation}
By using Eqs.~\eqref{IALD} and \eqref{11111LD}, we obtain
\begin{align}
	\left<T\right> &\approx \sum_{t^{\prime}=1}^{\infty} \frac{t^{\prime}}{2N} q\left(\frac{t^{\prime}}{2N}\right) \notag\\
	&= \frac{1}{2N} \frac{1}{2(2N-1)}
    \begin{pmatrix}
      0 & 2 & 0 & 0 & 2
    \end{pmatrix} (I - A^{{\rm LD}})^{-2} \bm{p}(0) \notag\\
  &= \frac{1}{2N} 
    \begin{pmatrix}
      1 & 1 & 1 & 1 & 1
    \end{pmatrix} (I - A^{{\rm LD}})^{-1} \bm{p}(0) \notag \\
  &= \frac{(2N-1)}{N(2N+3)} 
    \begin{pmatrix}
      N^2+2N+2 & \frac{2N^2+2N+1}{2} & \frac{4N^2+3N+3}{2} & \frac{4N^2+5N+6}{2} & \frac{4N^2-N+2}{2}
    \end{pmatrix} \bm{p}(0).
\end{align}
Because
\begin{equation}
	\left<T\right> = \begin{pmatrix}
      O(N) & O(N) & O(N) & O(N) & O(N)
    \end{pmatrix} \bm{p}(0), \label{dualTLD}
\end{equation}
we conclude $\left<T\right>=O(N)$ under LD.

\subsubsection{VM}\label{sub:double VM}

As shown in Appendix~\ref{sec:double-star VM},
the derivation of $\left<T\right>$ for the double-star graph under VM is similar to the case under LD. For this case, we obtain
\begin{equation}
	\left<T\right> =  \begin{pmatrix}
      O(1) & O(1) & O(N) & O(N) & O(N)
    \end{pmatrix}
     \bm{p}(0).
\label{eq:T double-star dual VM}
\end{equation}
We conclude $\left<T\right> = O(N)$
because $\left<T\right> = O(N)$ holds true for generic initial conditions corresponding to $p_3(0)$, $p_4(0)$, and $p_5(0)$.

\subsubsection{IP}\label{sub:double IP}

As shown in Appendix~\ref{sec:double-star IP},
$\left<T\right>$ for the double-star graph under IP is given by
\begin{equation}
	\left<T\right> =  \begin{pmatrix}
      O(N^2) & O(N^2) & O(N^3) & O(N^3) & O(N^3)
    \end{pmatrix}
     \bm{p}(0). 
\label{eq:T double-star dual IP}
\end{equation}
We conclude $\left<T\right> = O(N^3)$ under IP
because $\left<T\right> = O(N^3)$ holds true for generic initial conditions corresponding to $p_3(0)$, $p_4(0)$, and $p_5(0)$.

\section{Numerical simulations for large networks}\label{sec:direct numerical}

To check the validity of the scaling between $\left<T\right>$ and $N$ derived in Sec.~\ref{sec:order}, we carry out direct numerical simulations of the opinion dynamics for larger networks than those considered in Sec.~\ref{sec:exact}.
As the lollipop graph, we consider those having $N/2$ nodes in the chain and 
clique. As the barbell graph, we consider those having $N/3$ nodes in the chain and each clique. In each run,
$N/2$ randomly selected nodes initially possess state $\mathbf{0}$ and the other $N/2$ nodes state $\mathbf{1}$. We calculate $\left<T\right>$ as an average over 
$10^3$ runs for each network and update rule.

\begin{figure}
\begin{center}
\includegraphics[width=8cm]{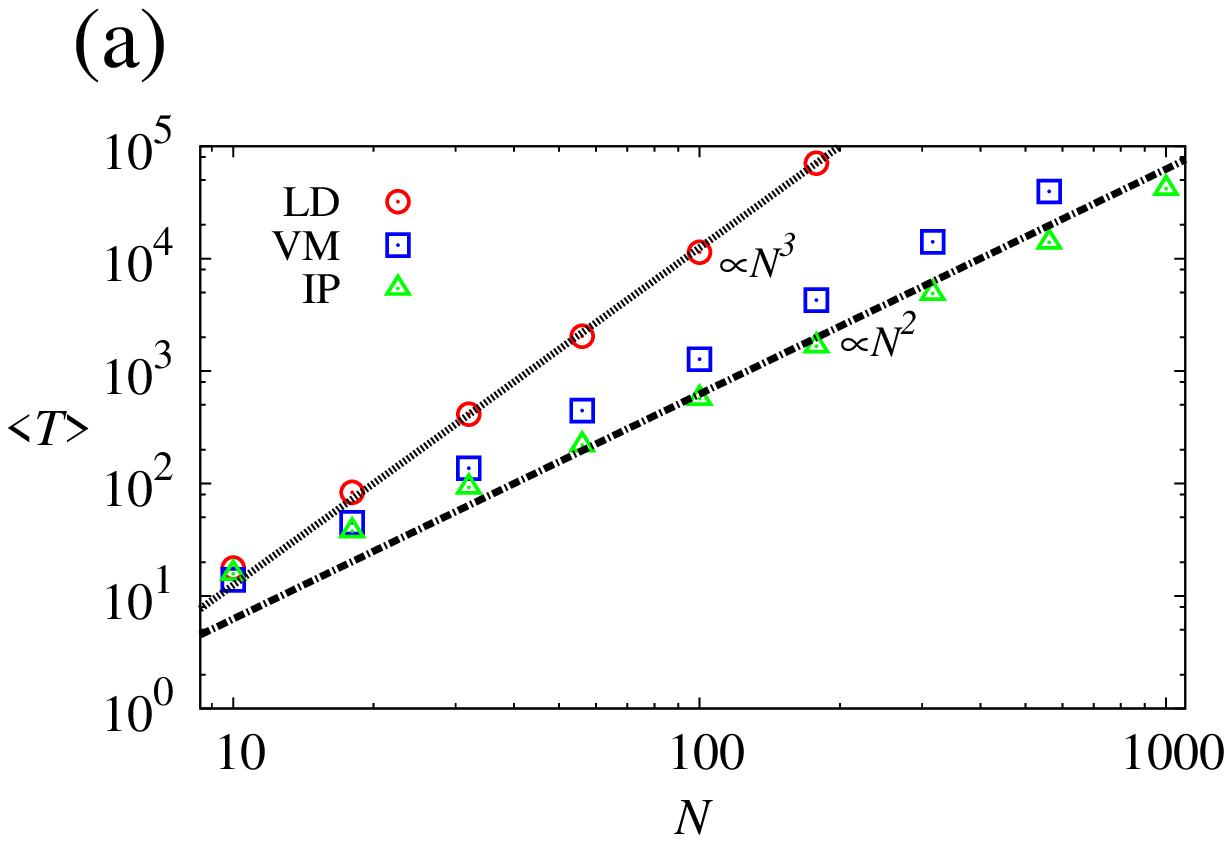}
\includegraphics[width=8cm]{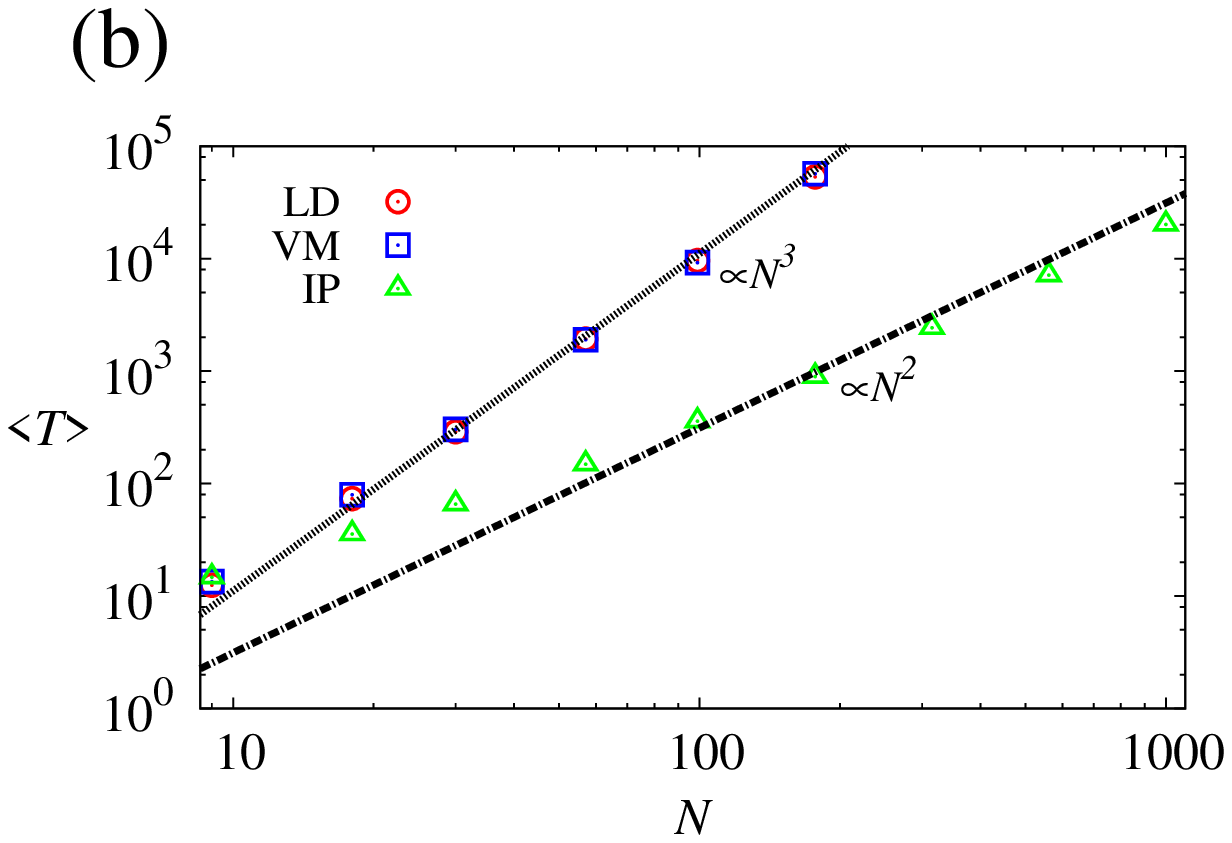}
\includegraphics[width=8cm]{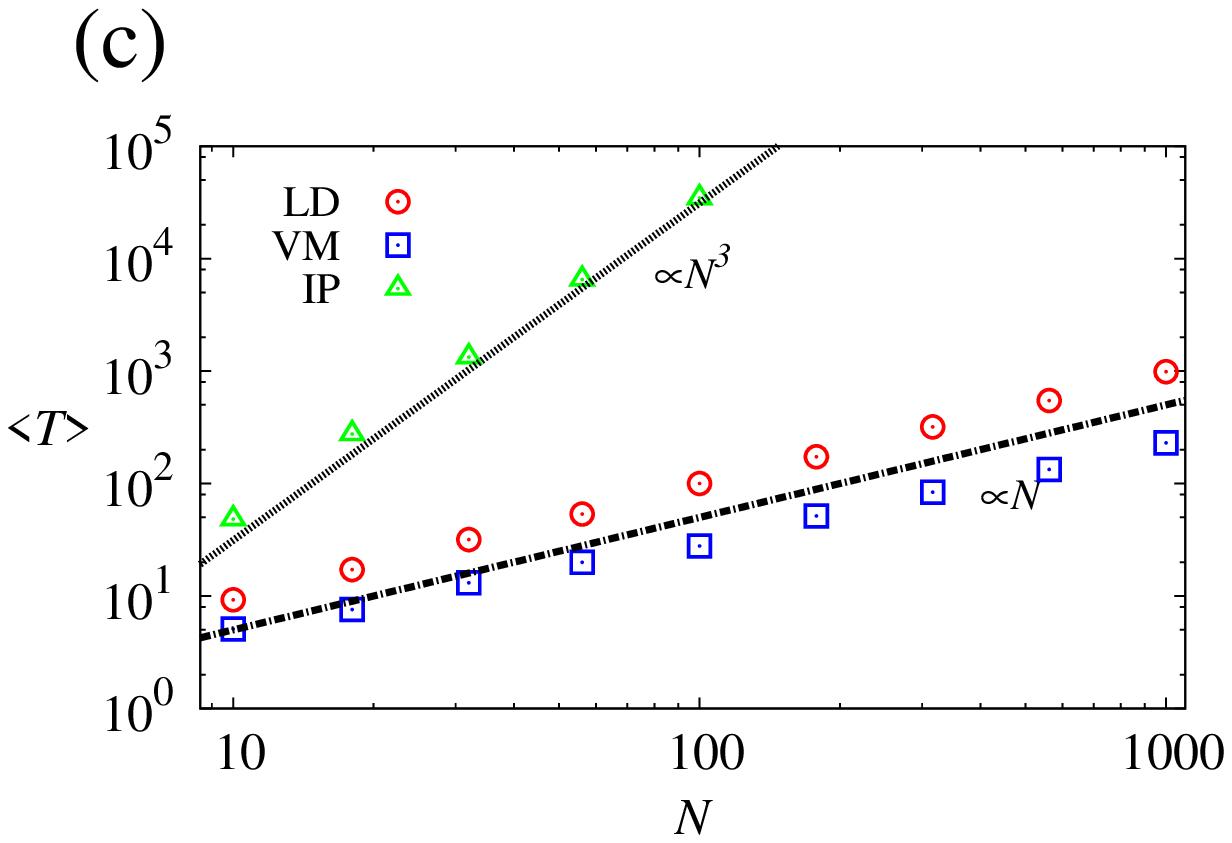}
\caption{Relationship between the mean consensus time, $\left<T\right>$, and the number of nodes, $N$, under different networks and update rules. (a) Lollipop graph. (b) Barbell graph. (c) Double-star graph.}
\label{fig:large N}
\end{center}
\end{figure}

The relationship between the numerically obtained $\left<T\right>$ and $N$
is shown in Fig.~\ref{fig:large N} for each combination of the network (i.e., lollipop, barbell, or double star) and update rule (i.e., LD, VM, or IP).
The numerical results (symbols) are largely consistent with
the scaling law derived in Sec.~\ref{sec:order} (lines) in most cases.

To be more quantitative, we fitted the relationship $\left<T\right>\propto N^{\alpha}$ to each plot shown in Fig.~\ref{fig:large N} using the least-square error method. For the double-star graph under VM, we added three data points ($\left<T\right>\approx 446.78$ for $N=1778$, $\left<T\right>\approx 781.06$ for $N=3162$, and $\left<T\right>\approx 1415.3$ for $N=5624$) to those shown in Fig.~\ref{fig:large N} before carrying out the least-square error method.
The numerically obtained $\alpha$ values, shown in Table~\ref{tab:alpha numerical}, are close to the theoretical
results summarized in Table~\ref{tab:summary} except for notable differences in some cases. For example, for the barbell graph under IP, the theory predicts $\alpha=2$, whereas the numerical results yield $\alpha\approx 1.81$. Although
 the precise reason for the discrepancy is unclear, it seems to be due to the finite size effect. For example, if we only use the data up to $N=1000$ for the double-star graph under VM, we would obtain
$\alpha\approx 0.88$. By extending the numerical simulations up to $N=5624$, we have obtained $\alpha\approx 1.00$; the theory predicts $\alpha=1$. In other combinations of network and update rule, we could not carry out numerical simulations for larger populations due to the computational cost.

\section{Discussion}

We explored the networks that maximized the mean consensus time, $\left<T\right>$, of the three variants of the voter model. The lollipop graph, barbell graph, and double-star graph were suggested to maximize $\left<T\right>$ under the LD, VM, and IP update rules, respectively. In addition, we evaluated $\left<T\right>$ for the three types of networks under each of the three update rules. The results are summarized in Table~\ref{tab:summary}.

Although the dual process of the opinion dynamics is the coalescing random walk \cite{Liggett1985book,Donnelly1983MPCPS,Durrett1988book,Cox1989AnnProb},
we expect that the characteristic time of the coalescing random walk, such as the time to the final coalesence, and that of usual random walks, such as the hitting time, are qualitatively the same. If we accept this contention, our
results are consistent with the previous results for
random walks. The hitting and cover time for the random walk on the lollipop graph \cite{Lawler1986DiscMath,Chandra1989STOC,Brightwell1990RandStructAlgor} and the barbell graph
\cite{Mazo1982BellSysTechJ,Coppersmith1993SiamJDiscMath}
both scale as $O(N^3)$. For the lollipop graph, the theoretical results \cite{Lawler1986DiscMath,Chandra1989STOC,Brightwell1990RandStructAlgor} are consistent with ours for LD. For the barbell graph, the theoretical results \cite{Mazo1982BellSysTechJ,Coppersmith1993SiamJDiscMath} are consistent with ours for LD and VM.
However, different update rules yield
$\left<T\right>=O(N^2)$
for the lollipop and barbell graphs.
The scaling between $\left<T\right>$ and $N$
depends on the update rule because a choice of the update rule corresponds to weighting of the links in the network. The link weight biases the probability that a particular link is used for state updating in opinion dynamics and the probability with which the random walk transits from one node to another \cite{Antal2006PRL,Sood2008PRE,MasudaOhtsuki2009NewJPhys}. It should be noted that more exact estimation of $\left<T\right>$
for the lollipop and barbell graphs on the basis of the random walk, as we did for the double-star graph, warrants future work.

The maximum hitting time of the random walk with respect to the network structure scales as $O(N^3)$ \cite{Lawler1986DiscMath,Brightwell1990RandStructAlgor,Coppersmith1993SiamJDiscMath}. Therefore, we expect that the consensus time
$O(N^3)$ attained for some combinations of the network and update rule in the present study is 
the maximum possible except for the constant factor and nonleading terms.

The double-star graph, which maximizes $\left<T\right>$ under IP, is 
far from the lollipop and barbell graphs in two aspects. First, it has a small diameter, i.e., three. Second, it has not been recognised as a network that slows down dynamics of the random walk. 
Previous theoretical results for opinion dynamics in
heterogeneous random networks yielded
$\left<T\right>=O(N \mu_1 \mu_{-1})$ for IP,
where $\mu_1$ and $\mu_{-1}$ are the mean degree and the mean of the inverse degree, respectively \cite{Antal2006PRL,Sood2008PRE}.
For the double star with $N$ nodes, where $N$ is even, this theory predicts 
$\left<T\right>=O(N)$ because
$\mu_1 = 2(N-1)/N$ and $\mu_{-1}=(N^2-2N+4)/N^2$. This estimate deviates from our results, i.e.,
$\left<T\right>=O(N^3)$. In the theory developed in 
Refs.~\cite{Antal2006PRL,Sood2008PRE}, heterogeneous random networks are assumed such that the network does not have structure other than the degree distribution. In contrast, in the double-star graph, leaf nodes are never adjacent to each other, and the two hubs are always adjacent. We consider this is the reason for the deviation. Similarly, the theory for heterogeneous random networks
\cite{Antal2006PRL,Sood2008PRE}
adapted to the degree distribution of the double-star graph suggests
$\left<T\right>=N\mu_1^2 / \mu_2 = O(1)$ under VM, where
$\mu_2=(N^2+2N-4)/2N$ is the second moment of the degree. This estimate is also different from ours, i.e., $\left<T\right>=O(N)$, presumably for the same reason.

\section*{Acknowledgments}

We thank Yusuke Kobayashi for suggesting a succinct proof for the upper bound of the mean consensus time under the LD update rule,
Sidney Redner for valuable discussion, and Ryosuke Nishi for careful reading of the manuscript.
N.M. acknowledges the support provided through JST, CREST, and JST, ERATO, Kawarabayashi Large Graph Project.

\appendix

\section{Derivation of $\left<T\right>$ for the double-star graph under VM}\label{sec:double-star VM}

On the double-star graph,
consider a single update event under VM. There are two types of events.
First, a leaf node imitates the
  state of the hub node in the same class with probability $1/2N$. In the
  random walk interpretation, this event corresponds to the movement
  of a walker located at this leaf (if so) to the hub. Second, a hub imitates the opinion of a neighbor, which is either
  a leaf in the same class or the hub in the opposite class, with
  probability $1/2N^2$. In the random walk, this event corresponds to
  the movement of a walker located at the hub to a
  neighbor. Neither walker moves if 
a walker is not located at the selected starting node. Therefore, we obtain
\begin{align}
  \bm{p}\left(t+\frac{1}{2N}\right) &= A^{{\rm VM}} \bm{p}(t),\\
  q\left(t+\frac{1}{2N}\right) &= \frac{1}{2N^2} 
    \begin{pmatrix}
      0 & N+1 & 0 & 0 & 2
    \end{pmatrix}
   \bm{p}(t),
\end{align}
where
\begin{equation}
	A^{{\rm VM}} = \frac{1}{2N^2} 
    \begin{pmatrix}
      2N(N-1) & N-2 & 0 & 0 & 0 \\
      2N & 2N(N-1) & 1 & 0 & 0 \\
      0 & 1 & 2N(N-1) & 2N & 2(N-1) \\
      0 & 0 & N-1 & 2N(N-1) & 0 \\
      0 & 0 & N & 0 & 2N(N-1)
    \end{pmatrix}. \label{AVM}
\end{equation}
Equation~\eqref{AVM} leads to
\begin{align}
	(I - A^{{\rm VM}})^{-1} = \frac{N^2}{2N+3} 
    \begin{pmatrix}
      \frac{4N-1}{N} & \frac{2(N-2)}{N} & \frac{N-2}{N} & \frac{N-2}{N} & \frac{(N-1)(N-2)}{N^2} \\[1.5ex]
      4 & 4 & 2 & 2 & \frac{2(N-1)}{N} \\[1.5ex]
      2 & 2 & 2(N+2) & 2(N+2) & \frac{2(N-1)(N+2)}{N} \\[1.5ex]
      \frac{N-1}{N} & \frac{N-1}{N} & \frac{(N-1)(N+2)}{N} & \frac{N^2+3N+1}{N} & \frac{(N-1)^2(N+2)}{N^2} \\[1.5ex]
      1 & 1 & N+2 & N+2 & \frac{N^2+3N+1}{N}
    \end{pmatrix} \label{IAVM}
\end{align}
and
\begin{equation}
	\frac{1}{2N^2} 
    \begin{pmatrix}
      0 & N+1 & 0 & 0 & 2
    \end{pmatrix}
   (I - A^{{\rm VM}})^{-1} \\
  = \begin{pmatrix}
      1 & 1 & 1 & 1 & 1
    \end{pmatrix}. \label{11111VM}
\end{equation}
By using Eqs.~\eqref{IAVM} and \eqref{11111VM}, we obtain
\begin{align}
	\left<T\right> &\approx \sum_{t^{\prime}=1}^{\infty} \frac{t^{\prime}}{2N} q\left(\frac{t^{\prime}}{2N}\right)\notag\\
  &= \frac{1}{2N(2N+3)} 
    \begin{pmatrix}
      a^{\rm VM} & b^{\rm VM} & c^{\rm VM} & d^{\rm VM} & e^{\rm VM}
    \end{pmatrix}
   \bm{p}(0),
\end{align}
where
\begin{align}
	a^{\rm VM} &= 2N(6N-1),\\
	b^{\rm VM} &= 5N(2N-1),\\
	c^{\rm VM} &= 2N(2N^2+5N-2),\\
	d^{\rm VM} &= N(4N^2+12N-1),\\
	e^{\rm VM} &= 4N^3+8N^2-11N+4.
\end{align}
Therefore, we obtain Eq.~\eqref{eq:T double-star dual VM}.

\section{Derivation of $\left<T\right>$ for the double-star graph under IP}\label{sec:double-star IP}

Under IP, a walker at a leaf of the double-star graph moves to the hub in the same class with probability $1/2N^2$ in an update event. A walker at a hub moves to a leaf in the same class with probability $1/2N$ and to the hub in the opposite class with probability $1/2N^2$. With the remaining probability, neither walker moves. Therefore, we obtain
\begin{align}
  \bm{p}\left(t+\frac{1}{2N}\right) &= A^{{\rm IP}} \bm{p}(t),\\
  q\left(t+\frac{1}{2N}\right) &= \frac{1}{2N^2} 
    \begin{pmatrix}
      0 & N+1 & 0 & 0 & 2
    \end{pmatrix}
   \bm{p}(t),
\end{align}
where
\begin{equation}
	A^{{\rm IP}} = \frac{1}{2N^2} 
    \begin{pmatrix}
      2(N+1)(N-1) & N(N-2) & 0 & 0 & 0 \\
      2 & (N+2)(N-1) & 1 & 0 & 0 \\
      0 & 1 & (N+2)(N-1) & 2 & 2N(N-1) \\
      0 & 0 & N(N-1) & 2(N+1)(N-1) & 0 \\
      0 & 0 & 1 & 0 & 2(N-1)
    \end{pmatrix}. \label{AIP}
\end{equation}
Equation~\eqref{AIP} leads to
\begin{align}
	&(I - A^{{\rm IP}})^{-1} = \frac{2N^2}{N^3+N+3} \times \notag \\ 
    &\begin{pmatrix}
      \frac{N^4-2N^3+4N^2-3N+3}{2} & \frac{(C +1)N(N-2)}{2} & \frac{C N(N-2)}{2} & \frac{C N(N-2)}{2} & \frac{N^2(N-1)(N-2)}{2} \\[1.5ex]
      C +1 & C +1 & C & C & C-1 \\[1.5ex]
      C & C & C (N+2) & C (N+2) & N(N-1)(N+2) \\[1.5ex]
      \frac{C N(N-1)}{2} & \frac{C N(N-1)}{2} & \frac{C N(N-1)(N+2)}{2} & \frac{N^5-N^3+3N^2-N+3}{2} & \frac{N^2(N-1)^2(N+2)}{2} \\[1.5ex]
      \frac{1}{2} & \frac{1}{2} & \frac{N+2}{2} & \frac{N+2}{2} & \frac{2N+3}{2}
    \end{pmatrix} \label{IAIP}
\end{align}
and
\begin{equation}
	\frac{1}{2N^2} 
    \begin{pmatrix}
      0 & N+1 & 0 & 0 & 2
    \end{pmatrix}
   (I - A^{{\rm IP}})^{-1} \\
  = \begin{pmatrix}
      1 & 1 & 1 & 1 & 1
    \end{pmatrix}, \label{11111IP}
\end{equation}
where $C \equiv N^2-N+1$. By using Eqs.~\eqref{IAIP} and \eqref{11111IP},
we obtain
\begin{align}
	\left<T\right> &\approx \sum_{t^{\prime}=1}^{\infty} \frac{t^{\prime}}{2N} q\left(\frac{t^{\prime}}{2N}\right)\notag\\
  &= \frac{N}{N^3+N+3} 
    \begin{pmatrix}
      a^{\rm IP} & b^{\rm IP} & c^{\rm IP} & d^{\rm IP} & e^{\rm IP}
    \end{pmatrix}
   \bm{p}(0),
\end{align}
where
\begin{align}
	a^{\rm IP} &= N^4-2N^3+5N^2-4N+5,\\
	b^{\rm IP} &= \frac{2N^4-5N^3+10N^2-9N+7}{2},\\
	c^{\rm IP} &= \frac{N^5+N^4-3N^3+10N^2-7N+8}{2},\\
	d^{\rm IP} &= \frac{N^5+N^4-2N^3+10N^2-6N+11}{2},\\
	e^{\rm IP} &= \frac{N^5+N^4-4N^3+8N^2-4N+3}{2}.
\end{align}
Therefore, we obtain Eq.~\eqref{eq:T double-star dual IP}.


\newpage
\clearpage

\clearpage

\clearpage

\clearpage

\clearpage

\clearpage

\clearpage

\clearpage

\includegraphics[width=10cm]{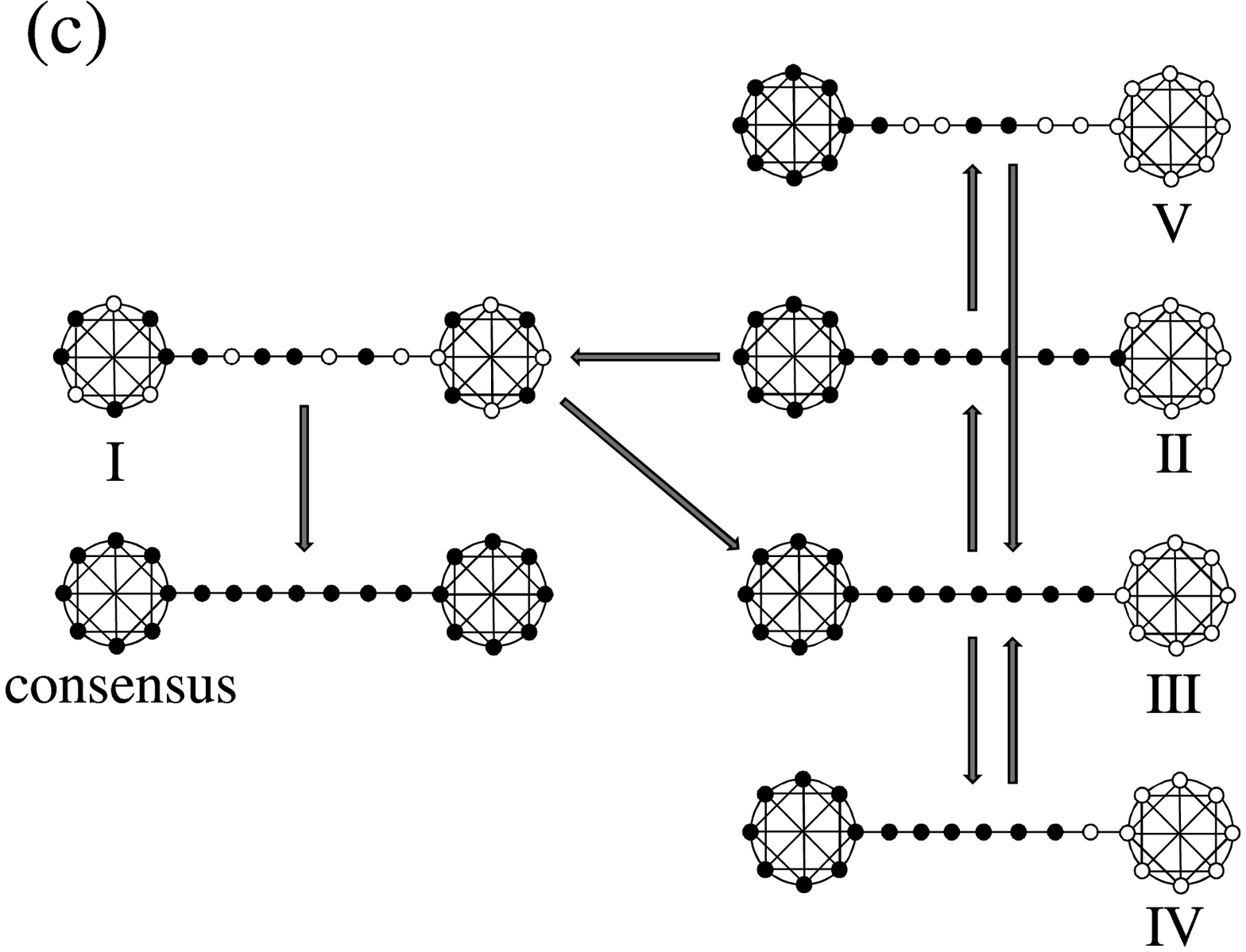}

\clearpage

\clearpage

\clearpage

\begin{table}[H]
\caption{Summary of the results. The dependence of the analytical estimations of mean consensus times on the number of nodes is shown for each combination of the network and update rule.}
\label{tab:summary}
\begin{center}
	\begin{tabular}{c|ccc}
		& LD & VM & IP \\ \hline
		Lollipop & $O(N^3)$ & $O(N^2)$ & $O(N^2)$ \\
		Barbell & $O(N^3)$ & $O(N^3)$ & $O(N^2)$ \\
		Double star & $O(N)$ & $O(N)$ & $O(N^3)$
	\end{tabular}
\end{center}
\end{table}

\clearpage

\begin{table}[H]
\caption{Power-law exponent $\alpha$ for $\left<T\right>\propto N^{\alpha}$ obtained from the least-square error method applied to the numerical results.}
\label{tab:alpha numerical}
\begin{center}
	\begin{tabular}{c|ccc}
		& LD & VM & IP \\ \hline
		Lollipop & 3.03 & 2.11 & 1.89\\
		Barbell & 2.94 & 3.12 & 1.81\\
		Double star & 1.00 & 1.00 & 2.90
	\end{tabular}
\end{center}
\end{table}

\end{document}